\documentclass[aps,prl,twocolumn,superscriptaddress]{revtex4-1}

\usepackage{amssymb}
\usepackage{natbib}
\usepackage{multirow}
\usepackage{bm}
\usepackage{amsmath}
\usepackage{graphicx}
\usepackage{epstopdf}
\usepackage{subfigure}
\usepackage{natbib}
\usepackage{epsfig}
\usepackage{amsfonts}
\usepackage{mathrsfs}
\usepackage{braket}
\usepackage{tabularx}
\usepackage{xcolor}
\usepackage[toc,page,title,titletoc,header]{appendix}

\usepackage{dsfont,amsthm,amsbsy}

\usepackage{lineno}
\begin{document}

\title{Magnetic Hofstadter cascade in a twisted semiconductor homobilayer}
\author{Benjamin A. Foutty}
\affiliation{Department of Physics, Stanford University, Stanford, CA 94305, USA}
\affiliation{Geballe Laboratory for Advanced Materials, Stanford, CA 94305, USA}

\author{Aidan P. Reddy}
\affiliation{Department of Physics, Massachusetts Institute of Technology, Cambridge, Massachusetts 02139, USA}

\author{Carlos R. Kometter}
\affiliation{Department of Physics, Stanford University, Stanford, CA 94305, USA}
\affiliation{Geballe Laboratory for Advanced Materials, Stanford, CA 94305, USA}

\author{Kenji Watanabe}
\affiliation{Research Center for Electronic and Optical Materials, National Institute for Materials Science, 1-1 Namiki, Tsukuba 305-0044, Japan}

\author{Takashi Taniguchi}
\affiliation{Research Center for Materials Nanoarchitectonics, National Institute for Materials Science, 1-1 Namiki, Tsukuba 305-0044, Japan}

\author{Trithep Devakul}
\affiliation{Department of Physics, Stanford University, Stanford, CA 94305, USA}

\author{Benjamin E. Feldman}
\email{bef@stanford.edu}
\affiliation{Department of Physics, Stanford University, Stanford, CA 94305, USA}
\affiliation{Geballe Laboratory for Advanced Materials, Stanford, CA 94305, USA}
\affiliation{Stanford Institute for Materials and Energy Sciences, SLAC National Accelerator Laboratory, Menlo Park, CA 94025, USA}

\begin{abstract}
Transition metal dichalcogenide moir\'e homobilayers have emerged as a platform in which magnetism, strong correlations, and topology are intertwined. In a large magnetic field, the energetic alignment of states with different spin in these systems is dictated by both strong Zeeman splitting and the structure of the Hofstadter's butterfly spectrum, yet the latter has been difficult to probe experimentally. Here we conduct local thermodynamic measurements of twisted WSe$_2$ homobilayers that reveal a cascade of magnetic phase transitions. We understand these transitions as the filling of individual Hofstadter subbands, allowing us to extract the structure and connectivity of the Hofstadter spectrum of a single spin. The onset of magnetic transitions is independent of twist angle, indicating that the exchange interactions of the component layers are only weakly modified by the moir\'e potential. In contrast, the magnetic transitions are associated with stark changes in the insulating states at commensurate filling. Our work achieves a spin-resolved measurement of Hofstadter’s butterfly despite overlapping states, and it disentangles the role of material and moiré effects on the nature of the correlated ground states. 
\end{abstract}

\maketitle

Understanding the interplay between strong electronic correlations, symmetry breaking, and topology is of both fundamental and technical interest. Hofstadter's butterfly, which arises at high magnetic fields due to interference between the magnetic length and the crystalline lattice, provides a natural venue to realize these ingredients. When the magnetic flux per unit cell reaches a significant fraction of the flux quantum, the electronic spectrum reconstructs into fractal topological bands, termed `Hofstadter subbands' \cite{hofstadter_energy_1976}. These subbands carry integer Chern numbers and have garnered theoretical interest for hosting novel electronic phenomena \cite{shaffer_unconventional_2022,mishra_effects_2016,hong_harpers_1999,moller_fractional_2015}. The large, tunable unit cells of moir\'e superlattices make this problem experimentally accessible with laboratory scale magnetic fields \cite{bistritzer_moire_2011,moon_energy_2012,dean_hofstadters_2013,hunt_massive_2013,ponomarenko_cloning_2013,spanton_observation_2018} and can simultaneously generate flat electronic bands with strong interaction effects. 

Among moir\'e systems, semiconducting transition metal dichalcogenide (TMD) homobilayers present a unique limit of Hofstadter behavior. The low energy TMD valence bands have a twofold degenerate spin-valley locked isospin degree of freedom, referred to as spin below \cite{xu_spin_2014}. In a magnetic field, spin and orbital effects lead to a large single-particle Zeeman energy, which is further enhanced by exchange interactions at low carrier densities \cite{gustafsson_ambipolar_2018,shi_odd-_2020,foutty_anomalous_2024}. Additionally, because the TMD moir\'e system lacks $C_{2z}$-symmetry, the Hofstadter spectra of the two spins evolve differently with field \cite{kolar_hofstadter_2024,wang_phase_2024,zhao_hofstadter_2024}. These characteristics contrast with magic-angle twisted bilayer graphene (and similar systems), in which the Hofstadter spectra are degenerate and interactions drive the formation of symmetry-broken integer and fractional Chern insulating gaps, similar to quantum Hall ferromagnetism \cite{saito_hofstadter_2021,nuckolls_strongly_2020,wu_chern_2021,park_flavour_2021,das_symmetry-broken_2021,yu_correlated_2022}. However, probing the Hofstadter spectrum itself and how it is affected by interactions remains experimentally challenging, and full understanding of the TMD platform demands a new spin-selective approach.

\begin{figure*}[t!]
    \renewcommand{\thefigure}{\arabic{figure}}
    \centering
    \includegraphics[scale =1.0]{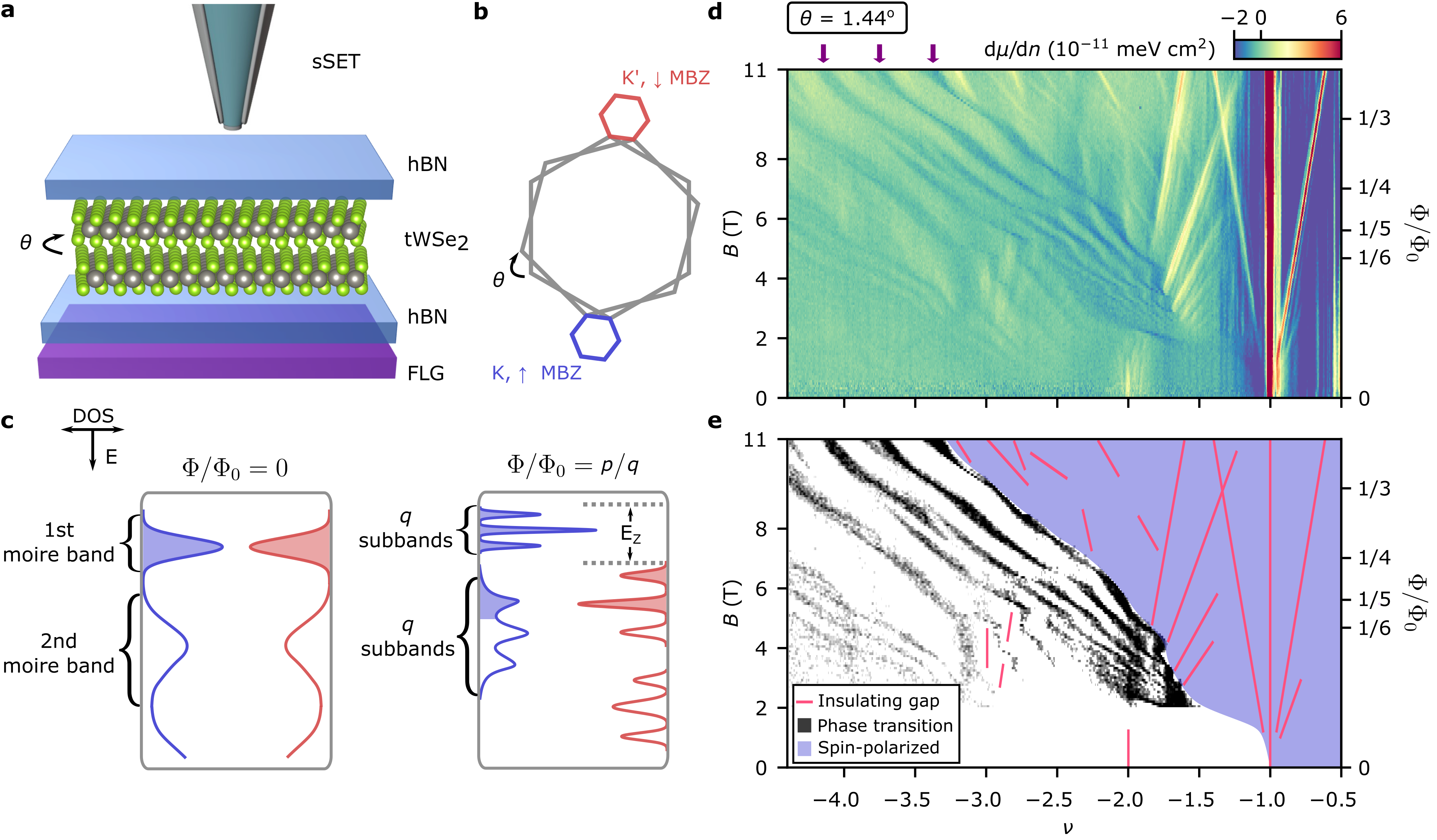}
    \caption{\textbf{Cascade of electronic phase transitions in twisted WSe$_2$ (tWSe$_2$).} \textbf{a}, Schematic of the scanning single-electron transistor (sSET) and device. \textbf{b}, Momentum space structure of tWSe$_2$: spin-valley locked moir\'e bands are localized in mini Brillouin zones (MBZs) near the $K$ and $K'$ points. \textbf{c}, Schematic of the electronic density of states (DOS) of tWSe$_2$ valence bands for each spin at both zero magnetic flux quanta per unit cell $\Phi / \Phi_0$ and at a large, rational value $\Phi/\Phi_0 = p/q$, where $\Phi_0$ is the flux quantum, and $p,q$ are integers. At nonzero flux, the Zeeman energy $E_Z$ shifts the valence band edges and each moir\'e band splits into $q$ Hofstadter subbands. \textbf{d}, Inverse electronic compressibility d$\mu$/d$n$ at twist angle $\theta = 1.44^\circ$ as a function of moir\'e filling factor $\nu$ and magnetic field $B$ (left axis) or equivalently $\Phi/\Phi_0$ (right axis). \textbf{e}, Wannier style plot schematically showing incompressible gaps, the phase transitions from \textbf{d}, and the region of full spin polarization.}
    \label{fig:Fig1}
\end{figure*}

The ordering of microscopic degrees of freedom is also relevant to the numerous correlated ground states realized in TMD homobilayers \cite{mak_semiconductor_2022,wang_correlated_2020,ghiotto_quantum_2021}. For example, integer and fractional quantum anomalous Hall states in these systems rely on ferromagnetic spin polarization \cite{wu_topological_2019,devakul_magic_2021,anderson_programming_2023,foutty_mapping_2024,cai_signatures_2023,zeng_thermodynamic_2023,park_observation_2023,xu_observation_2023}. The layer degree of freedom also plays a key role, with layer hybridization and polarization giving rise to topological and trivial bands respectively. Additionally, recently discovered unconventional superconductivity depends strongly on layer-polarizing electric fields  \cite{guo_superconductivity_2024,xia_unconventional_2024}. It is therefore desirable to characterize how distinct experimental tuning knobs couple to and affect the energetic competition between different broken-symmetry ground states. A related demand is discerning the role of the moir\'e superlattice itself in driving symmetry breaking, versus when underlying material properties are dominant \cite{zhou_half-_2021}. Addressing these questions will inform efforts to deterministically pursue particular topological and correlated electronic behavior of interest.

In this work, we use a scanning single electron transistor (SET) to probe the electronic compressibility and magnetization of twisted bilayer WSe$_2$ (tWSe$_2$). We observe a cascade of transitions as Hofstadter subbands of distinct spins are sequentially populated. These phase transitions reveal the spin-resolved structure of the Hofstadter spectrum. Through comparison of the behavior at different twist angles, we show that the onset of magnetic transitions is only weakly influenced by the moir\'e superlattice. In contrast, the phase transitions couple strongly to adjacent insulating states driven by moir\'e physics. Collectively, our measurements demonstrate that the underlying exchange interactions from the constituent TMD determine the sample magnetization in the small twist angle limit, which can in turn dictate the stability and nature of the correlated ground states.

\begin{figure*}[t!]
    \renewcommand{\thefigure}{\arabic{figure}}
    \centering
    \includegraphics[scale =1.0]{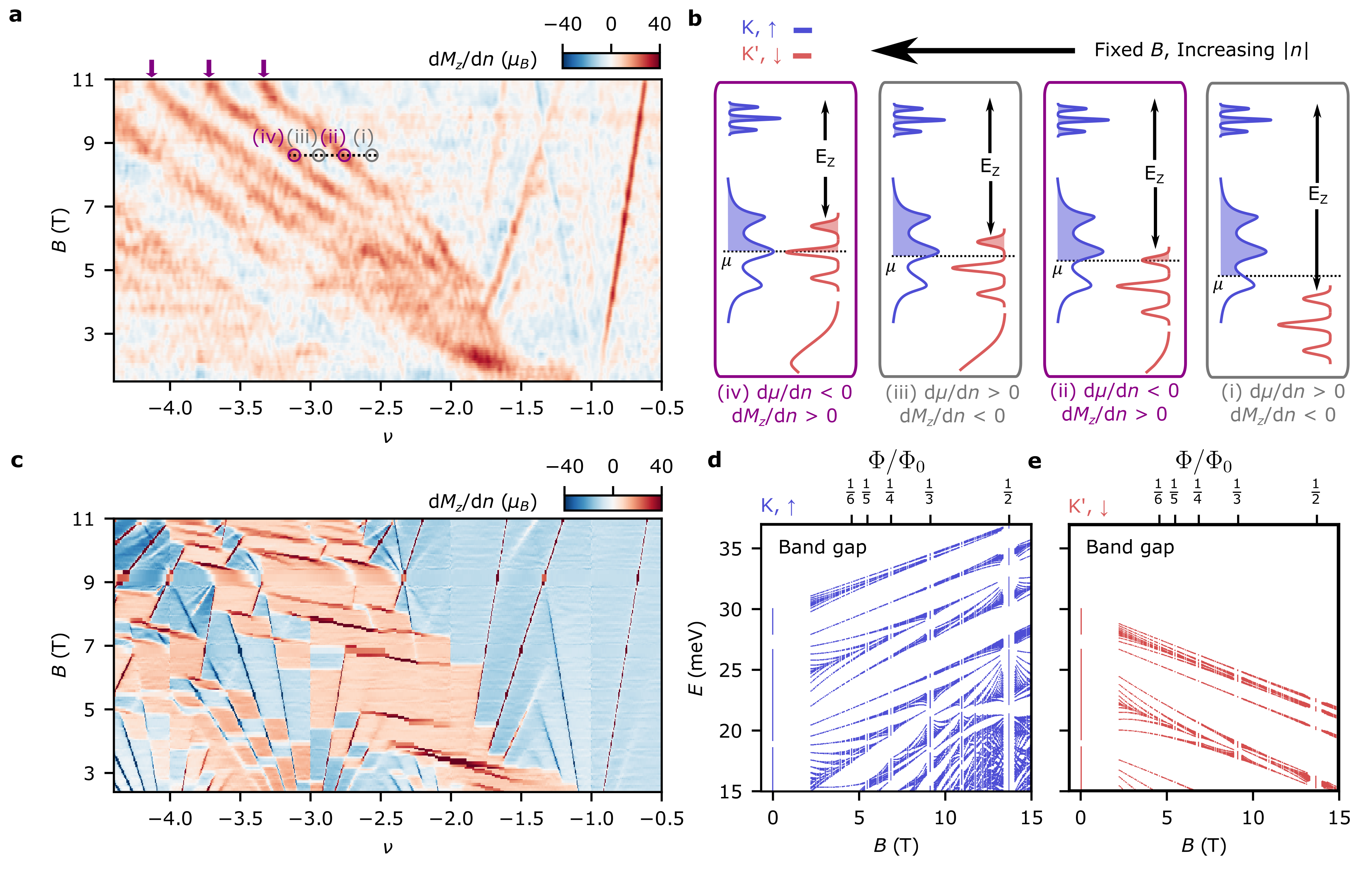}
    \caption{\textbf{Magnetization of phase transitions and Hofstadter subband crossings.} \textbf{a}, Derivative of the magnetization in the $z$ direction, d$M_z$/d$n$, as a function of $\nu$ and $B$. \textbf{b}, Schematic of Hofstadter subband crossings as carrier density $n$ is changed at fixed field. As $|n|$ is increased, individual spin-minority ($K',\downarrow$) Hofstadter subbands cross the Fermi level, denoted by $\mu$.
    \textbf{c}, Calculated d$M_z$/d$n$ from a Stoner model using the Hofstadter spectra shown in \textbf{d}-\textbf{e} (Methods).
    \textbf{d}-\textbf{e}, Calculated spin-resolved Hofstadter spectra based on the continuum model tWSe$_2$ bands (Methods). 
    }
    \label{fig:Fig2}
\end{figure*}

Figure \ref{fig:Fig1}a depicts the experimental setup. We study a tWSe$_2$ device with a local twist angle $\theta$ that varies between $1.1^\circ$ and $1.6^\circ$ in different areas of the sample, with uniform domains much larger than the SET tip (Methods). The low-energy electronic states arise from moir\'e valence bands of the two spin flavors, which are degenerate at zero magnetic field (Fig.~\ref{fig:Fig1}b-c). In a large magnetic field, the Zeeman energy $E_Z$ lifts this degeneracy by an amount that can significantly exceed the moir\'e bandwidths \cite{zhao_realization_2024,kometter_hofstadter_2023,ghiotto_stoner_2024,park_ferromagnetism_2024}, and the moir\'e bands are further split into subbands due to Hofstadter physics (Fig.~\ref{fig:Fig1}c).

\begin{figure*}[t!]
    \renewcommand{\thefigure}{\arabic{figure}}
    \centering
    \includegraphics[scale =1.0]{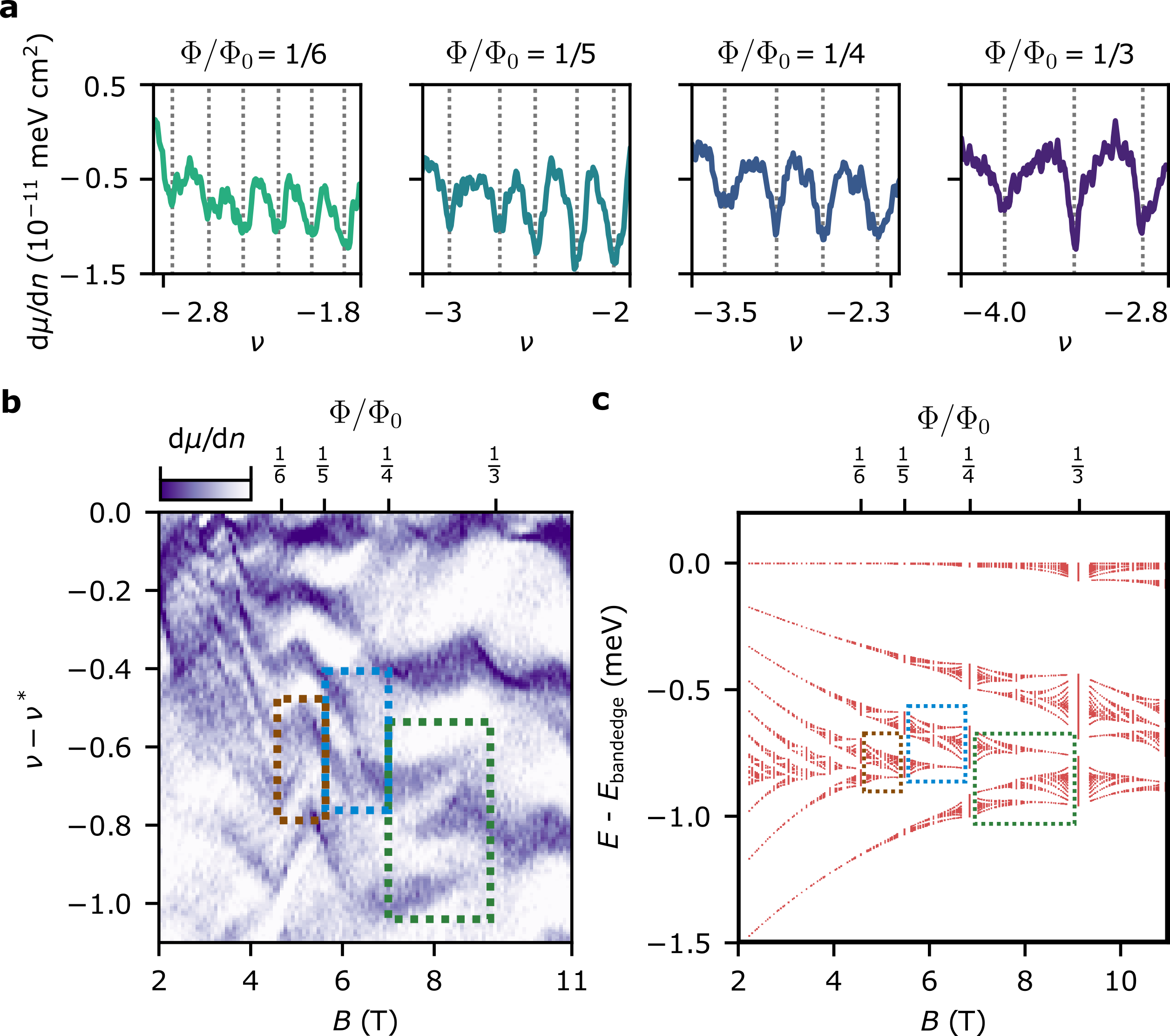}
    \caption{\textbf{Determining the Hofstadter spectrum structure.} \textbf{a}, d$\mu$/d$n$ as a function of $\nu$ for a set of rational magnetic flux quanta per unit cell $\Phi/\Phi_0$, showing $q$ dips when $\Phi/\Phi_0 = 1/q$. The data are taken from Fig.~\ref{fig:Fig1}d. \textbf{b}, d$\mu$/d$n$ as a function of $B$ and $\nu - \nu^*$, the difference in filling factor relative to the onset of magnetic transitions. The span of the colorbar is from $-1 \times 10^{-11}$ meV cm$^{2}$ to 0. Dashed boxes highlight how subbands merge going from $\Phi/\Phi_0 = \frac{1}{6} \rightarrow \frac{1}{5}$ (brown), $\Phi/\Phi_0 = \frac{1}{5} \rightarrow \frac{1}{4}$ (blue) and $\Phi/\Phi_0 = \frac{1}{4} \rightarrow \frac{1}{3}$ (green). \textbf{c}, Calculated spin-minority Hofstadter spectrum, plotted in energy relative to the valence band edge $E - E_{\rm{bandedge}}$ and magnetic field. Dashed boxes highlight regions that can be identified in \textbf{b}.}
    \label{fig:Fig3}
\end{figure*}

In Fig.~\ref{fig:Fig1}d, we show the inverse electronic compressibility d$\mu$/d$n$ measured as a function of moir\'e filling factor $\nu$ and out-of-plane magnetic field $B$ in a location with $\theta = 1.44^\circ$. The field axis can also be expressed in terms of the number of flux quanta per moir\'e unit cell $\Phi/\Phi_0$, where $\Phi_0 = h/e$ is the flux quantum, $h$ is Planck's constant, and $e$ is the charge of the electron. We observe multiple incompressible states, including insulators which persist to $B = 0$ (at $\nu = -1,-2$) and Hofstadter states emerging at nonzero magnetic fields. Each insulating gap is described by the Diophantine equation $\nu = t(\Phi/\Phi_0) + s$, where $t$ is the Chern number and $s$ is the intercept at zero magnetic field (pink lines in Fig.~\ref{fig:Fig1}e) \cite{streda_theory_1982}. These gaps are similar to those observed in previous reports in TMD systems \cite{kometter_hofstadter_2023,foutty_mapping_2024}.

We also resolve a prominent series of dips in d$\mu$/d$n$ (e.g., purple arrows in Fig.~\ref{fig:Fig1}d) that traverse the $\nu$-$B$ plane. Negative electronic compressibility of this form is strongly indicative of isospin phase transitions between distinct electronic states \cite{eisenstein_negative_1992,feldman_fractional_2013,zondiner_cascade_2020,zhou_half-_2021,yu_correlated_2022}. To understand this cascade of transitions, we consider expectations for the spin polarization in different regimes. At low hole densities and high magnetic fields, the Zeeman effect completely polarizes the system into a single spin favored by the Zeeman energy (`spin-majority', pale blue shading in Fig.~\ref{fig:Fig1}e) \cite{kometter_hofstadter_2023,park_ferromagnetism_2024}. As carriers are added or the magnetic field is decreased, opposite spin (`spin-minority') states begin to be filled, suggesting that the transitions may be related to changes in spin occupation. This motivates examination of the magnetic behavior across the transitions.

From our measurement of the chemical potential $\mu$ as a function of $n$ and $B$, we can use a Maxwell relation to determine the magnetization $M_z$ (Methods). Its derivative with respect to carrier density, d$M_z$/d$n$ (Fig.~\ref{fig:Fig2}a) reveals that both the incompressible Hofstadter states and the cascade of phase transitions carry significant changes in $M_z$. In the case of the Hofstadter states (for example, $(t,s) = (+1,-1)$ appears red in Fig.~\ref{fig:Fig2}a), this can be understood from the finite orbital magnetization of states with nonzero Chern number. In particular, the orbital magnetization jumps by $\Delta M_z = t \frac{e}{h} \Delta \mu$ across a Hofstadter gap of size $\Delta \mu$ \cite{zhu_voltage-controlled_2020,zeng_thermodynamic_2023,redekop_direct_2024}. The steps in magnetization from the spin transitions can be classified into two characteristic behaviors. At low magnetic fields, the single step (around $\nu = -1.8$ at $B = 2$ T) can be understood as a moir\'e band crossing, which carries a change in $M_z$ of roughly 10 $\mu_B$ per moir\'e unit cell (Extended Data Fig.~\ref{fig:MagLinetraces}) \cite{kometter_hofstadter_2023, foutty_mapping_2024, park_ferromagnetism_2024}. At higher fields, this band crossing splits into a series of discrete steps, but the total change of magnetization is similar (Extended Data Fig.~\ref{fig:MagLinetraces}).

The cascade of magnetization transitions is caused by changes in the occupation of spin-minority Hofstadter subbands as they cross the Fermi level. We illustrate this schematically in Fig.~\ref{fig:Fig2}b. As the first spin-minority subband crosses the Fermi level and begins to fill with holes, the total magnetization decreases and d$M_z$/d$n > 0$ (Fig.~\ref{fig:Fig2}b[ii],[iv]). When including Stoner magnetic interactions which disfavor occupation of carriers of both spin flavors, this coincides with a depletion of spin-majority holes, leading to a negative d$\mu$/d$n$ (Supplementary Sec. 1-2) \cite{zondiner_cascade_2020,kometter_hofstadter_2023}. Eventually the spin-minority Hofstadter subband is completely filled (Fig.~\ref{fig:Fig2}b[iii]). At this point, the Fermi level lies in a gap between spin-minority subbands, and spin-majority states are again filled with additional holes, returning the derivatives of magnetization and chemical potential to their previous values. We therefore identify a 1-to-1 correspondence between the pattern of magnetic transitions and the spin-minority Hofstadter spectrum: regions of negative d$\mu/$d$n$ correspond to filling subbands, and the spacing between these regions corresponds to gaps in the spectrum.

\begin{figure}[t!]
    \renewcommand{\thefigure}{\arabic{figure}}
    \centering
    \includegraphics[scale =1.0]{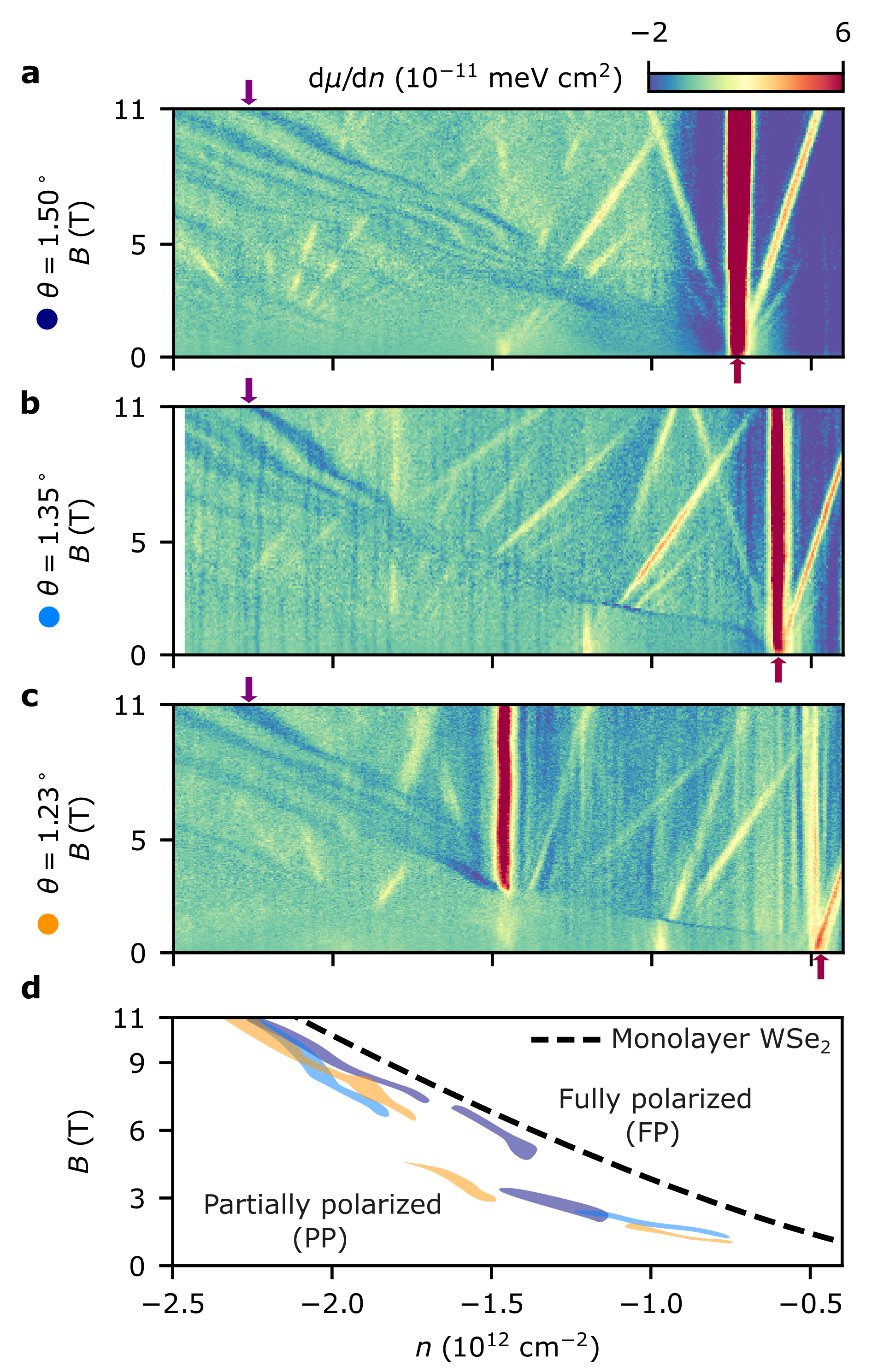}
    \caption{\textbf{Magnetic phase transitions at distinct twist angles.} \textbf{a}-\textbf{c}, d$\mu$/d$n$ as a function of $n$ and $B$ at $\theta = 1.50^\circ$ (\textbf{a}), $\theta = 1.35^\circ$ (\textbf{b}), and $\theta = 1.23^\circ$ (\textbf{c}). Maroon arrows highlight correlated insulating gaps at $\nu=-1$, which occur at distinct carrier densities due to the changing twist angle. Purple arrows highlight magnetic transitions with negative d$\mu$/d$n$ at high magnetic fields, which appear at roughly the same carrier density independent of twist angle.
    \textbf{d}, Schematic showing the onset of magnetic phase transitions (at lowest hole density) for each twist angle from \textbf{a}-\textbf{c}, as well as the approximate location of the transition between full and partial spin polarization in monolayer WSe$_2$ for comparison.}
    \label{fig:Fig4}
\end{figure}

We can directly compare the observed d$M_z$/d$n$ to the predictions of a Stoner model using the Hofstadter spectra of the calculated tWSe$_2$ continuum model band structure, shown in Fig.~\ref{fig:Fig2}c-e (Methods). This simulation captures a number of features of the experimental data, including the sharp single transition at low magnetic fields, in which the entire first moir\'e band crosses the Fermi level, and the splitting of transitions in the Hofstadter regime at higher magnetic fields. One notable divergence between theory and experiment is that we observe better separated phase transitions (pink regions in Fig \ref{fig:Fig2}a,c) in the experimental data, which suggests that the spin-minority Hofstadter spectrum has well separated gaps and that the spin-majority density of states may have weaker gaps than predicted 
by our Stoner model beyond the first moir\'e band. We discuss possible reasons for this in Supplementary Sec. 3.

The number of transitions we observe at different fields further confirms their association with the Hofstadter spectrum.  The number of Hofstadter subbands is related to the number of magnetic flux quanta threading through each moir\'e unit cell: when $\Phi/\Phi_0 = p/q$ for integer $p,q$, each moir\'e band splits into $q$ subbands \cite{hofstadter_energy_1976}. In Fig.~\ref{fig:Fig3}a, we plot d$\mu$/d$n$ at several magnetic fields corresponding to $\Phi/\Phi_0 = 1/q$ for $q = 3,4,5,6$, focusing on the series of phase transitions near the onset of partial spin polarization stemming from the first moir\'e band crossing at low magnetic fields. For this set of rational fluxes, we observe $q$ dips at each flux, consistent with the expectations from the Hofstadter spectrum. 

By examining the behavior of the phase transitions between these rational fluxes, we can also establish the connectivity and structure of the spin-minority Hofstadter spectrum. In Fig.~\ref{fig:Fig3}b, we show the full magnetic field dependence of d$\mu$/d$n$ for this set of phase transitions, now plotted as moir\'e filling relative to the onset of phase transitions, $\nu^*$. The evolution of the subbands that we observe in experiment matches the theoretical calculations of the spin-minority Hofstadter spectrum, shown in Fig.~\ref{fig:Fig3}c. For example, the third and fourth subbands at $\Phi/\Phi_0 = \frac{1}{5}$ (counting from the valence band edge) combine into a single subband at $\Phi/\Phi_0 = \frac{1}{4}$, with the third subband connecting smoothly and the fourth subband vanishing (blue dashed box in Fig.~\ref{fig:Fig3}b-c). Between $\Phi/\Phi_0 = \frac{1}{4}$ and $\frac{1}{3}$, the shape is qualitatively different: the third and fourth subbands smoothly recede between these magnetic fields, while a new subband grows between them, becoming the third subband from the valence band edge at $\frac{1}{3}$ (green dashed box in Fig.~\ref{fig:Fig3}b-c). We note that there is a subsequent transition at higher hole densities 
than the set shown in Fig.~\ref{fig:Fig3}a-b. We attribute this feature, and subsequent features, to the Landau level pinned to the edge of the band (and higher moir\'e bands), and provide a more complete discussion in Supplementary Sec. 3. More broadly, our technique can be viewed as conducting spectroscopy of the spin-minority Hofstadter spectrum with the spin-majority states.

\begin{figure*}[t!]
    \renewcommand{\thefigure}{\arabic{figure}}
    \centering
    \includegraphics[scale =1.0]{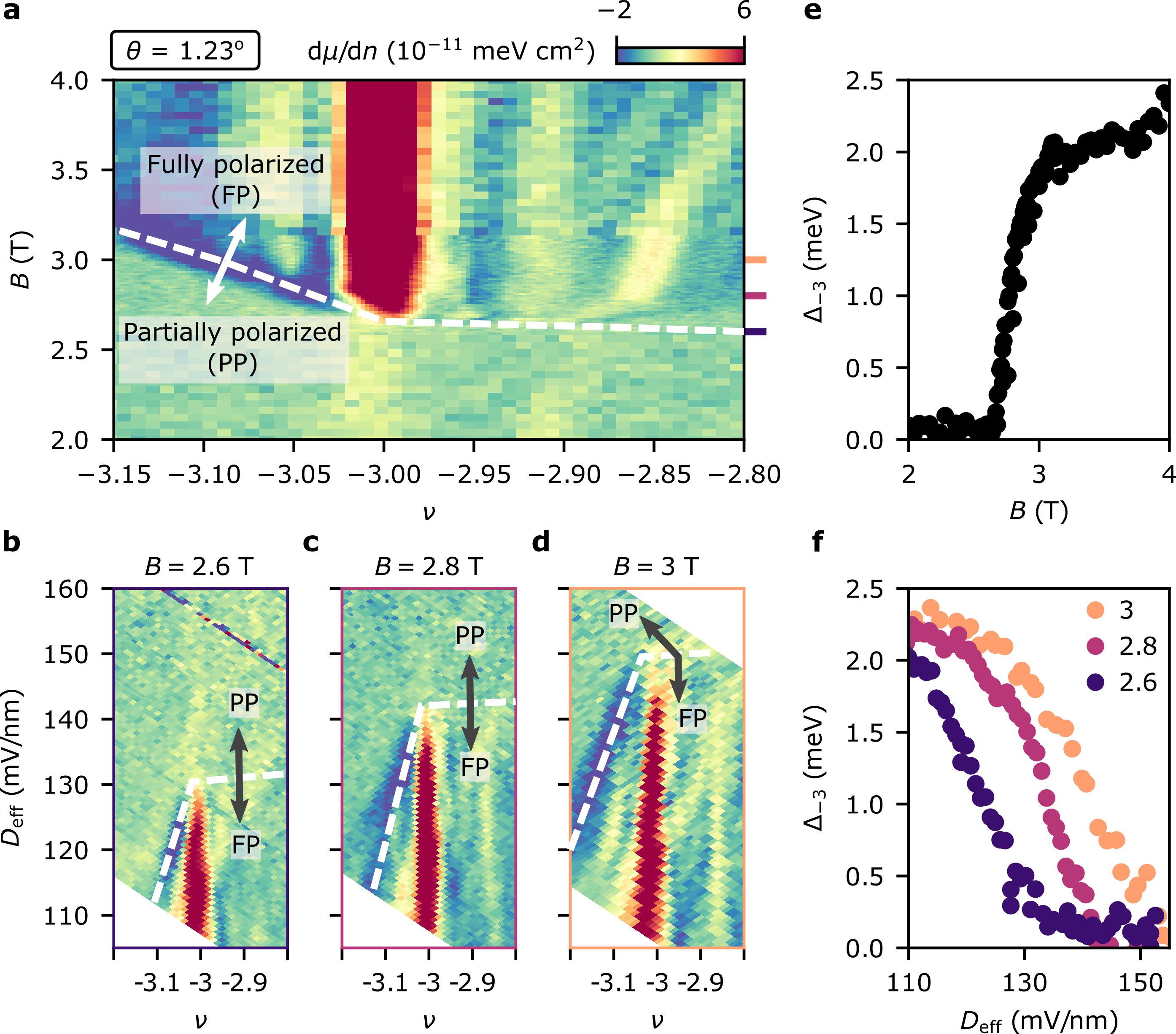}
    \caption{\textbf{Coupled spin and layer degrees of freedom near integer moir\'e filling.} \textbf{a}, d$\mu$/d$n$ as a function of $\nu$ and $B$ in the vicinity of the magnetic phase transition near $\nu = -3$ at a location with $\theta = 1.23^\circ$. The white dashed line is a guide to eye highlighting the phase transition between full spin polarization (FP) and partial polarization (PP). These data are taken in the absence of tip doping, with an effective displacement field $D_{\rm{eff}} \approx 140$ mV/nm. \textbf{b}-\textbf{d}, d$\mu$/d$n$ as a function of $\nu$ and $D_{\rm{eff}}$ at fixed magnetic fields. The fully spin-polarized state is favored at lower $D_{\rm{eff}}$, and both $B$ and $D_{\rm{eff}}$ affect the location of the metal insulator transition at $\nu = -3$. \textbf{e}-\textbf{f}, Thermodynamic gap $\Delta_{-3}$ at $\nu = -3$ as a function of $B$ with no tip doping (from numerically integrating across the incompressible peak in \textbf{a}) and as a function of $D_{\rm{eff}}$ at various $B$ (from \textbf{b}-\textbf{d}).}
    \label{fig:Fig5} 
\end{figure*}

In the above discussion, we have focused on the behavior at a single location within the device. As the underlying electronic structure and correlated states are strongly affected by the local twist angle \cite{wang_correlated_2020,foutty_mapping_2024,knuppel_correlated_2024}, a natural question is the extent to which twist angle also modifies the magnetic transitions. To address this, we compare d$\mu$/d$n$ at multiple locations with different twist angles, plotted as a function of the absolute carrier density $n$ and $B$ (Fig.~\ref{fig:Fig4}a-c). The correlated insulators and Hofstadter states occur at varying $n$ due to the changing size of the moir\'e unit cell. However, the magnetic phase transitions occur at similar locations within the $n$-$B$ plane for all of the observed twist angles between $\theta = 1.1^\circ$ and $\theta = 1.6^\circ$ (Fig.~\ref{fig:Fig4}d, see Extended Data Fig.~\ref{fig:OtherTwistsdmudn} and Supplementary Sec. 4 for data from additional angles). This behavior stands in contrast to measurements of tWSe$_2$ at higher twist angles, in which van Hove singularities in the moir\'e bands have a strong effect on where polarization occurs in the electronic phase diagram \cite{ghiotto_stoner_2024,knuppel_correlated_2024,guo_superconductivity_2024}. The lack of dependence on $\theta$ at small twist angles, across which there are significant changes in moir\'e band structure \cite{devakul_magic_2021,zhang_polarization-driven_2024}, shows that a naive approach of comparing the Zeeman energy and bandwidth is insufficient to account for the onset of spin-polarization.

Instead, our data suggest that the underlying properties of the constituent materials are the dominant factor determining where spin transitions occur. Even in monolayer WSe$_2$, interaction effects are relatively strong due to large effective mass and low dielectric constant \cite{wang_valley-_2017,gustafsson_ambipolar_2018,shi_odd-_2020,shi_bilayer_2022,foutty_anomalous_2024}. The transition from full to partial spin polarization in monolayer WSe$_2$ (black dashed line in Fig.~\ref{fig:Fig4}d) follows a similar curve in $n$-$B$ space that roughly matches the onset of polarization transitions in tWSe$_2$. We postulate that the spin polarization in this system is inherited from and largely dictated by the underlying exchange physics within monolayer WSe$_2$. In this picture, the effect of the moir\'e superlattice is mainly to promote the formation of insulating gaps at both zero and finite magnetic fields. This reflects a broader trend in moir\'e systems, in which phase transitions in materials with strong interaction effects are only moderately adjusted by the addition of a moir\'e superlattice potential \cite{zhou_half-_2021}. 

While the magnetic transitions are not strongly affected by the moir\'e superlattice across much of the $\nu$-$B$ plane, they do couple strongly to the insulating phases at integer filling of the moir\'e bands. In Fig.~\ref{fig:Fig5}, we focus on the compressibility near the magnetic phase transition at $\nu = -3$ in a location with twist angle $\theta = 1.23^\circ$. At this filling, the magnetic phase transition is concurrent with a sharp metal-insulator transition in the correlated insulating state (Fig.~\ref{fig:Fig5}a). We also observe similar metal-insulator transitions or prominent changes in the strength of insulators at other twist angles (Extended Data Figs.~\ref{fig:OtherTwistsdmudn},\ref{fig:Nu=-3_OtherTwists}). Further, at numerous integer fillings beyond $\nu = -1$ (Extended Data Fig.~\ref{fig:DFieldDep_OtherTransitions}, Fig.~\ref{fig:Fig4}), we see changes in the pattern of incompressible Hofstadter states across the magnetic transitions, consistent with a reordering of the set of filled moir\'e bands and/or subbands. 

The moir\'e electronic states in homobilayers can generally be tuned by a perpendicular electric displacement field \cite{wang_correlated_2020,ghiotto_quantum_2021,cai_signatures_2023,zeng_thermodynamic_2023}. By changing the relative voltage between tip and sample, we can adjust the local displacement field $D_{\rm{eff}}$ within a finite range to examine how it affects the magnetic transitions (Methods). At most positions in the $\nu$-$B$ plane, changing $D_{\rm{eff}}$ does not noticeably shift the position of the magnetic transitions (Extended Data Fig.~\ref{fig:DFieldDep_OtherTransitions}). This indicates that, broadly, the layer polarization of the spin-minority Hofstadter subbands does not differ significantly from that of the competing spin-majority bands at the Fermi level. However, when the transitions coincide with changes in the moir\'e-driven correlated insulators, $D_{\rm{eff}}$ has a large effect on both the metal-insulator transition and the adjacent negative compressibility. By fixing the magnetic field, we can tune across the phase transitions by sweeping $D_{\rm{eff}}$, fully opening or closing the insulating gaps (Fig.~\ref{fig:Fig5}b-f). The fully spin polarized (high magnetic field) phase is favored by a lower displacement field, with similar dependence at $\nu=-2$ (Extended Data Fig.~\ref{fig:DFieldDep_OtherTransitions}). This likely stems from distinct microscopic layer polarizations and/or susceptibilities of the correlated phases on each side of the transition. One possible candidate for the fully spin-polarized insulating state at $\nu=-3$ and large $B$ is one in which the orbitals at each of the three high-symmetry stacking regions of the moiré unit cell are all filled \cite{qiu_interaction-driven_2023}. A full understanding of the microscopic nature of the low-$B$ states is beyond the scope of this work. However, the data demonstrate that at integer fillings, the moir\'e superlattice \emph{does} affect the position of the magnetic transitions, despite its negligible effects at higher carrier densities and non-integer fillings.

In conclusion, we have observed a cascade of magnetic transitions at high fields that are driven by the combination of large single-particle Zeeman energy and exchange interactions inherited from the parent WSe$_2$ material. Our measurements identify a novel manifestation of Hofstadter quantization beyond the formation of gaps. Specifically, we have observed structure in \emph{compressible} regions of the $\nu$-$B$ plane emerging from the interplay between Hofstadter quantization and electron-electron interactions. This provides a path to access flavor-specific Hofstadter spectra in a wide range of moir\'e platforms, which is otherwise difficult to measure in materials with competing degrees of freedom. Additionally, these measurements clarify how twist angle, carrier density, and out-of-plane electromagnetic fields can tune the spin polarization of the resulting ground states in tWSe$_2$, which can in turn inform rational materials design to stabilize desired topological and electronic properties.


\begin{thebibliography}{10}
\expandafter\ifx\csname url\endcsname\relax
  \def\url#1{\texttt{#1}}\fi
\expandafter\ifx\csname urlprefix\endcsname\relax\def\urlprefix{URL }\fi
\providecommand{\bibinfo}[2]{#2}
\providecommand{\eprint}[2][]{\url{#2}}

\bibitem{hofstadter_energy_1976}
\bibinfo{author}{Hofstadter, D.~R.}
\newblock \bibinfo{title}{Energy levels and wave functions of {Bloch} electrons in rational and irrational magnetic fields}.
\newblock \emph{\bibinfo{journal}{Physical Review B}} \textbf{\bibinfo{volume}{14}}, \bibinfo{pages}{2239--2249} (\bibinfo{year}{1976}).
\newblock \urlprefix\url{https://link.aps.org/doi/10.1103/PhysRevB.14.2239}.
\newblock \bibinfo{note}{Publisher: American Physical Society}.

\bibitem{shaffer_unconventional_2022}
\bibinfo{author}{Shaffer, D.}, \bibinfo{author}{Wang, J.} \& \bibinfo{author}{Santos, L.~H.}
\newblock \bibinfo{title}{Unconventional self-similar {Hofstadter} superconductivity from repulsive interactions}.
\newblock \emph{\bibinfo{journal}{Nature Communications}} \textbf{\bibinfo{volume}{13}}, \bibinfo{pages}{7785} (\bibinfo{year}{2022}).
\newblock \urlprefix\url{https://www.nature.com/articles/s41467-022-35316-z}.
\newblock \bibinfo{note}{Publisher: Nature Publishing Group}.

\bibitem{mishra_effects_2016}
\bibinfo{author}{Mishra, A.}, \bibinfo{author}{Hassan, S.~R.} \& \bibinfo{author}{Shankar, R.}
\newblock \bibinfo{title}{Effects of interaction in the {Hofstadter} regime of the honeycomb lattice}.
\newblock \emph{\bibinfo{journal}{Physical Review B}} \textbf{\bibinfo{volume}{93}}, \bibinfo{pages}{125134} (\bibinfo{year}{2016}).
\newblock \urlprefix\url{https://link.aps.org/doi/10.1103/PhysRevB.93.125134}.

\bibitem{hong_harpers_1999}
\bibinfo{author}{Hong, S.-P.} \& \bibinfo{author}{Suck~Salk, S.-H.}
\newblock \bibinfo{title}{Harper's equation for two-dimensional systems of antiferromagnetically correlated electrons}.
\newblock \emph{\bibinfo{journal}{Physical Review B}} \textbf{\bibinfo{volume}{60}}, \bibinfo{pages}{9550--9554} (\bibinfo{year}{1999}).
\newblock \urlprefix\url{https://link.aps.org/doi/10.1103/PhysRevB.60.9550}.
\newblock \bibinfo{note}{Publisher: American Physical Society}.

\bibitem{moller_fractional_2015}
\bibinfo{author}{Möller, G.} \& \bibinfo{author}{Cooper, N.~R.}
\newblock \bibinfo{title}{Fractional {Chern} {Insulators} in {Harper}-{Hofstadter} {Bands} with {Higher} {Chern} {Number}}.
\newblock \emph{\bibinfo{journal}{Physical Review Letters}} \textbf{\bibinfo{volume}{115}}, \bibinfo{pages}{126401} (\bibinfo{year}{2015}).
\newblock \urlprefix\url{https://link.aps.org/doi/10.1103/PhysRevLett.115.126401}.
\newblock \bibinfo{note}{Publisher: American Physical Society}.

\bibitem{bistritzer_moire_2011}
\bibinfo{author}{Bistritzer, R.} \& \bibinfo{author}{MacDonald, A.~H.}
\newblock \bibinfo{title}{Moir{\textbackslash}'e butterflies in twisted bilayer graphene}.
\newblock \emph{\bibinfo{journal}{Physical Review B}} \textbf{\bibinfo{volume}{84}}, \bibinfo{pages}{035440} (\bibinfo{year}{2011}).
\newblock \urlprefix\url{https://link.aps.org/doi/10.1103/PhysRevB.84.035440}.
\newblock \bibinfo{note}{Publisher: American Physical Society}.

\bibitem{moon_energy_2012}
\bibinfo{author}{Moon, P.} \& \bibinfo{author}{Koshino, M.}
\newblock \bibinfo{title}{Energy spectrum and quantum {Hall} effect in twisted bilayer graphene}.
\newblock \emph{\bibinfo{journal}{Physical Review B}} \textbf{\bibinfo{volume}{85}}, \bibinfo{pages}{195458} (\bibinfo{year}{2012}).
\newblock \urlprefix\url{https://link.aps.org/doi/10.1103/PhysRevB.85.195458}.
\newblock \bibinfo{note}{Publisher: American Physical Society}.

\bibitem{dean_hofstadters_2013}
\bibinfo{author}{Dean, C.~R.} \emph{et~al.}
\newblock \bibinfo{title}{Hofstadter’s butterfly and the fractal quantum {Hall} effect in moiré superlattices}.
\newblock \emph{\bibinfo{journal}{Nature}} \textbf{\bibinfo{volume}{497}}, \bibinfo{pages}{598--602} (\bibinfo{year}{2013}).
\newblock \urlprefix\url{https://www.nature.com/articles/nature12186}.
\newblock \bibinfo{note}{Number: 7451 Publisher: Nature Publishing Group}.

\bibitem{hunt_massive_2013}
\bibinfo{author}{Hunt, B.} \emph{et~al.}
\newblock \bibinfo{title}{Massive {Dirac} {Fermions} and {Hofstadter} {Butterfly} in a van der {Waals} {Heterostructure}}.
\newblock \emph{\bibinfo{journal}{Science}} \textbf{\bibinfo{volume}{340}}, \bibinfo{pages}{1427--1430} (\bibinfo{year}{2013}).
\newblock \urlprefix\url{https://www.science.org/doi/10.1126/science.1237240}.
\newblock \bibinfo{note}{Publisher: American Association for the Advancement of Science}.

\bibitem{ponomarenko_cloning_2013}
\bibinfo{author}{Ponomarenko, L.~A.} \emph{et~al.}
\newblock \bibinfo{title}{Cloning of {Dirac} fermions in graphene superlattices}.
\newblock \emph{\bibinfo{journal}{Nature}} \textbf{\bibinfo{volume}{497}}, \bibinfo{pages}{594--597} (\bibinfo{year}{2013}).
\newblock \urlprefix\url{https://www.nature.com/articles/nature12187}.
\newblock \bibinfo{note}{Publisher: Nature Publishing Group}.

\bibitem{spanton_observation_2018}
\bibinfo{author}{Spanton, E.~M.} \emph{et~al.}
\newblock \bibinfo{title}{Observation of fractional {Chern} insulators in a van der {Waals} heterostructure}.
\newblock \emph{\bibinfo{journal}{Science}} \textbf{\bibinfo{volume}{360}}, \bibinfo{pages}{62--66} (\bibinfo{year}{2018}).
\newblock \urlprefix\url{https://www.science.org/doi/full/10.1126/science.aan8458}.
\newblock \bibinfo{note}{Publisher: American Association for the Advancement of Science}.

\bibitem{xu_spin_2014}
\bibinfo{author}{Xu, X.}, \bibinfo{author}{Yao, W.}, \bibinfo{author}{Xiao, D.} \& \bibinfo{author}{Heinz, T.~F.}
\newblock \bibinfo{title}{Spin and pseudospins in layered transition metal dichalcogenides}.
\newblock \emph{\bibinfo{journal}{Nature Physics}} \textbf{\bibinfo{volume}{10}}, \bibinfo{pages}{343--350} (\bibinfo{year}{2014}).
\newblock \urlprefix\url{https://www.nature.com/articles/nphys2942}.
\newblock \bibinfo{note}{Number: 5 Publisher: Nature Publishing Group}.

\bibitem{gustafsson_ambipolar_2018}
\bibinfo{author}{Gustafsson, M.~V.} \emph{et~al.}
\newblock \bibinfo{title}{Ambipolar {Landau} levels and strong band-selective carrier interactions in monolayer {WSe2}}.
\newblock \emph{\bibinfo{journal}{Nature Materials}} \textbf{\bibinfo{volume}{17}}, \bibinfo{pages}{411--415} (\bibinfo{year}{2018}).
\newblock \urlprefix\url{https://www.nature.com/articles/s41563-018-0036-2}.

\bibitem{shi_odd-_2020}
\bibinfo{author}{Shi, Q.} \emph{et~al.}
\newblock \bibinfo{title}{Odd- and even-denominator fractional quantum {Hall} states in monolayer {WSe2}}.
\newblock \emph{\bibinfo{journal}{Nature Nanotechnology}} \textbf{\bibinfo{volume}{15}}, \bibinfo{pages}{569--573} (\bibinfo{year}{2020}).
\newblock \urlprefix\url{https://www.nature.com/articles/s41565-020-0685-6}.
\newblock \bibinfo{note}{Number: 7 Publisher: Nature Publishing Group}.

\bibitem{foutty_anomalous_2024}
\bibinfo{author}{Foutty, B.~A.} \emph{et~al.}
\newblock \bibinfo{title}{Anomalous {Landau} {Level} {Gaps} {Near} {Magnetic} {Transitions} in {Monolayer} \$\{{\textbackslash}mathrm\{{WSe}\}\}\_\{2\}\$}.
\newblock \emph{\bibinfo{journal}{Physical Review X}} \textbf{\bibinfo{volume}{14}}, \bibinfo{pages}{031018} (\bibinfo{year}{2024}).
\newblock \urlprefix\url{https://link.aps.org/doi/10.1103/PhysRevX.14.031018}.
\newblock \bibinfo{note}{Publisher: American Physical Society}.

\bibitem{kolar_hofstadter_2024}
\bibinfo{author}{Kolář, K.}, \bibinfo{author}{Yang, K.}, \bibinfo{author}{von Oppen, F.} \& \bibinfo{author}{Mora, C.}
\newblock \bibinfo{title}{Hofstadter spectrum of {Chern} bands in twisted transition metal dichalcogenides} (\bibinfo{year}{2024}).
\newblock \urlprefix\url{http://arxiv.org/abs/2406.06680}.
\newblock \bibinfo{note}{ArXiv:2406.06680 [cond-mat]}.

\bibitem{wang_phase_2024}
\bibinfo{author}{Wang, M.}, \bibinfo{author}{Wang, X.} \& \bibinfo{author}{Vafek, O.}
\newblock \bibinfo{title}{Phase diagram of twisted bilayer \$\{{\textbackslash}mathrm\{{MoTe}\}\}\_\{2\}\$ in a magnetic field with an account for the electron-electron interaction}.
\newblock \emph{\bibinfo{journal}{Physical Review B}} \textbf{\bibinfo{volume}{110}}, \bibinfo{pages}{L201107} (\bibinfo{year}{2024}).
\newblock \urlprefix\url{https://link.aps.org/doi/10.1103/PhysRevB.110.L201107}.
\newblock \bibinfo{note}{Publisher: American Physical Society}.

\bibitem{zhao_hofstadter_2024}
\bibinfo{author}{Zhao, C.} \emph{et~al.}
\newblock \bibinfo{title}{Hofstadter spectrum in a semiconductor moiré lattice} (\bibinfo{year}{2024}).
\newblock \urlprefix\url{http://arxiv.org/abs/2406.08044}.
\newblock \bibinfo{note}{ArXiv:2406.08044 [cond-mat]}.

\bibitem{saito_hofstadter_2021}
\bibinfo{author}{Saito, Y.} \emph{et~al.}
\newblock \bibinfo{title}{Hofstadter subband ferromagnetism and symmetry-broken {Chern} insulators in twisted bilayer graphene}.
\newblock \emph{\bibinfo{journal}{Nature Physics}} \textbf{\bibinfo{volume}{17}}, \bibinfo{pages}{478--481} (\bibinfo{year}{2021}).
\newblock \urlprefix\url{https://www.nature.com/articles/s41567-020-01129-4}.
\newblock \bibinfo{note}{Publisher: Nature Publishing Group}.

\bibitem{nuckolls_strongly_2020}
\bibinfo{author}{Nuckolls, K.~P.} \emph{et~al.}
\newblock \bibinfo{title}{Strongly correlated {Chern} insulators in magic-angle twisted bilayer graphene}.
\newblock \emph{\bibinfo{journal}{Nature}} \textbf{\bibinfo{volume}{588}}, \bibinfo{pages}{610--615} (\bibinfo{year}{2020}).
\newblock \urlprefix\url{https://www.nature.com/articles/s41586-020-3028-8}.
\newblock \bibinfo{note}{Publisher: Nature Publishing Group}.

\bibitem{wu_chern_2021}
\bibinfo{author}{Wu, S.}, \bibinfo{author}{Zhang, Z.}, \bibinfo{author}{Watanabe, K.}, \bibinfo{author}{Taniguchi, T.} \& \bibinfo{author}{Andrei, E.~Y.}
\newblock \bibinfo{title}{Chern insulators, van {Hove} singularities and topological flat bands in magic-angle twisted bilayer graphene}.
\newblock \emph{\bibinfo{journal}{Nature Materials}} \textbf{\bibinfo{volume}{20}}, \bibinfo{pages}{488--494} (\bibinfo{year}{2021}).
\newblock \urlprefix\url{https://www.nature.com/articles/s41563-020-00911-2}.
\newblock \bibinfo{note}{Publisher: Nature Publishing Group}.

\bibitem{park_flavour_2021}
\bibinfo{author}{Park, J.~M.}, \bibinfo{author}{Cao, Y.}, \bibinfo{author}{Watanabe, K.}, \bibinfo{author}{Taniguchi, T.} \& \bibinfo{author}{Jarillo-Herrero, P.}
\newblock \bibinfo{title}{Flavour {Hund}’s coupling, {Chern} gaps and charge diffusivity in moiré graphene}.
\newblock \emph{\bibinfo{journal}{Nature}} \textbf{\bibinfo{volume}{592}}, \bibinfo{pages}{43--48} (\bibinfo{year}{2021}).
\newblock \urlprefix\url{https://www.nature.com/articles/s41586-021-03366-w}.
\newblock \bibinfo{note}{Number: 7852 Publisher: Nature Publishing Group}.

\bibitem{das_symmetry-broken_2021}
\bibinfo{author}{Das, I.} \emph{et~al.}
\newblock \bibinfo{title}{Symmetry-broken {Chern} insulators and {Rashba}-like {Landau}-level crossings in magic-angle bilayer graphene}.
\newblock \emph{\bibinfo{journal}{Nature Physics}} \textbf{\bibinfo{volume}{17}}, \bibinfo{pages}{710--714} (\bibinfo{year}{2021}).
\newblock \urlprefix\url{https://www.nature.com/articles/s41567-021-01186-3}.
\newblock \bibinfo{note}{Publisher: Nature Publishing Group}.

\bibitem{yu_correlated_2022}
\bibinfo{author}{Yu, J.} \emph{et~al.}
\newblock \bibinfo{title}{Correlated {Hofstadter} spectrum and flavour phase diagram in magic-angle twisted bilayer graphene}.
\newblock \emph{\bibinfo{journal}{Nature Physics}} \textbf{\bibinfo{volume}{18}}, \bibinfo{pages}{825--831} (\bibinfo{year}{2022}).
\newblock \urlprefix\url{https://www.nature.com/articles/s41567-022-01589-w}.
\newblock \bibinfo{note}{Number: 7 Publisher: Nature Publishing Group}.

\bibitem{mak_semiconductor_2022}
\bibinfo{author}{Mak, K.~F.} \& \bibinfo{author}{Shan, J.}
\newblock \bibinfo{title}{Semiconductor moiré materials}.
\newblock \emph{\bibinfo{journal}{Nature Nanotechnology}} \textbf{\bibinfo{volume}{17}}, \bibinfo{pages}{686--695} (\bibinfo{year}{2022}).
\newblock \urlprefix\url{https://www.nature.com/articles/s41565-022-01165-6}.
\newblock \bibinfo{note}{Number: 7 Publisher: Nature Publishing Group}.

\bibitem{wang_correlated_2020}
\bibinfo{author}{Wang, L.} \emph{et~al.}
\newblock \bibinfo{title}{Correlated electronic phases in twisted bilayer transition metal dichalcogenides}.
\newblock \emph{\bibinfo{journal}{Nature Materials}} \textbf{\bibinfo{volume}{19}}, \bibinfo{pages}{861--866} (\bibinfo{year}{2020}).
\newblock \urlprefix\url{https://www.nature.com/articles/s41563-020-0708-6}.

\bibitem{ghiotto_quantum_2021}
\bibinfo{author}{Ghiotto, A.} \emph{et~al.}
\newblock \bibinfo{title}{Quantum criticality in twisted transition metal dichalcogenides}.
\newblock \emph{\bibinfo{journal}{Nature}} \textbf{\bibinfo{volume}{597}}, \bibinfo{pages}{345--349} (\bibinfo{year}{2021}).
\newblock \urlprefix\url{https://www.nature.com/articles/s41586-021-03815-6}.

\bibitem{wu_topological_2019}
\bibinfo{author}{Wu, F.}, \bibinfo{author}{Lovorn, T.}, \bibinfo{author}{Tutuc, E.}, \bibinfo{author}{Martin, I.} \& \bibinfo{author}{MacDonald, A.}
\newblock \bibinfo{title}{Topological {Insulators} in {Twisted} {Transition} {Metal} {Dichalcogenide} {Homobilayers}}.
\newblock \emph{\bibinfo{journal}{Physical Review Letters}} \textbf{\bibinfo{volume}{122}}, \bibinfo{pages}{086402} (\bibinfo{year}{2019}).
\newblock \urlprefix\url{https://link.aps.org/doi/10.1103/PhysRevLett.122.086402}.
\newblock \bibinfo{note}{Publisher: American Physical Society}.

\bibitem{devakul_magic_2021}
\bibinfo{author}{Devakul, T.}, \bibinfo{author}{Crépel, V.}, \bibinfo{author}{Zhang, Y.} \& \bibinfo{author}{Fu, L.}
\newblock \bibinfo{title}{Magic in twisted transition metal dichalcogenide bilayers}.
\newblock \emph{\bibinfo{journal}{Nature Communications}} \textbf{\bibinfo{volume}{12}}, \bibinfo{pages}{6730} (\bibinfo{year}{2021}).
\newblock \urlprefix\url{https://www.nature.com/articles/s41467-021-27042-9}.
\newblock \bibinfo{note}{Number: 1 Publisher: Nature Publishing Group}.

\bibitem{anderson_programming_2023}
\bibinfo{author}{Anderson, E.} \emph{et~al.}
\newblock \bibinfo{title}{Programming correlated magnetic states with gate-controlled moiré geometry}.
\newblock \emph{\bibinfo{journal}{Science}} \textbf{\bibinfo{volume}{381}}, \bibinfo{pages}{325--330} (\bibinfo{year}{2023}).
\newblock \urlprefix\url{https://www.science.org/doi/10.1126/science.adg4268}.
\newblock \bibinfo{note}{Publisher: American Association for the Advancement of Science}.

\bibitem{foutty_mapping_2024}
\bibinfo{author}{Foutty, B.~A.} \emph{et~al.}
\newblock \bibinfo{title}{Mapping twist-tuned multiband topology in bilayer {WSe2}}.
\newblock \emph{\bibinfo{journal}{Science}} \textbf{\bibinfo{volume}{384}}, \bibinfo{pages}{343--347} (\bibinfo{year}{2024}).
\newblock \urlprefix\url{https://www.science.org/doi/full/10.1126/science.adi4728}.
\newblock \bibinfo{note}{Publisher: American Association for the Advancement of Science}.

\bibitem{cai_signatures_2023}
\bibinfo{author}{Cai, J.} \emph{et~al.}
\newblock \bibinfo{title}{Signatures of {Fractional} {Quantum} {Anomalous} {Hall} {States} in {Twisted} {MoTe2}}.
\newblock \emph{\bibinfo{journal}{Nature}} \bibinfo{pages}{1--3} (\bibinfo{year}{2023}).
\newblock \urlprefix\url{https://www.nature.com/articles/s41586-023-06289-w}.
\newblock \bibinfo{note}{Publisher: Nature Publishing Group}.

\bibitem{zeng_thermodynamic_2023}
\bibinfo{author}{Zeng, Y.} \emph{et~al.}
\newblock \bibinfo{title}{Thermodynamic evidence of fractional {Chern} insulator in moiré {MoTe2}}.
\newblock \emph{\bibinfo{journal}{Nature}} \bibinfo{pages}{1--2} (\bibinfo{year}{2023}).
\newblock \urlprefix\url{https://www.nature.com/articles/s41586-023-06452-3}.
\newblock \bibinfo{note}{Publisher: Nature Publishing Group}.

\bibitem{park_observation_2023}
\bibinfo{author}{Park, H.} \emph{et~al.}
\newblock \bibinfo{title}{Observation of {Fractionally} {Quantized} {Anomalous} {Hall} {Effect}}.
\newblock \emph{\bibinfo{journal}{Nature}} \bibinfo{pages}{1--3} (\bibinfo{year}{2023}).
\newblock \urlprefix\url{https://www.nature.com/articles/s41586-023-06536-0}.
\newblock \bibinfo{note}{Publisher: Nature Publishing Group}.

\bibitem{xu_observation_2023}
\bibinfo{author}{Xu, F.} \emph{et~al.}
\newblock \bibinfo{title}{Observation of integer and fractional quantum anomalous {Hall} effects in twisted bilayer {MoTe2}} (\bibinfo{year}{2023}).
\newblock \urlprefix\url{http://arxiv.org/abs/2308.06177}.
\newblock \bibinfo{note}{ArXiv:2308.06177 [cond-mat]}.

\bibitem{guo_superconductivity_2024}
\bibinfo{author}{Guo, Y.} \emph{et~al.}
\newblock \bibinfo{title}{Superconductivity in twisted bilayer {WSe}\$\_2\$} (\bibinfo{year}{2024}).
\newblock \urlprefix\url{http://arxiv.org/abs/2406.03418}.
\newblock \bibinfo{note}{ArXiv:2406.03418 [cond-mat]}.

\bibitem{xia_unconventional_2024}
\bibinfo{author}{Xia, Y.} \emph{et~al.}
\newblock \bibinfo{title}{Unconventional superconductivity in twisted bilayer {WSe2}} (\bibinfo{year}{2024}).
\newblock \urlprefix\url{http://arxiv.org/abs/2405.14784}.
\newblock \bibinfo{note}{ArXiv:2405.14784 [cond-mat]}.

\bibitem{zhou_half-_2021}
\bibinfo{author}{Zhou, H.} \emph{et~al.}
\newblock \bibinfo{title}{Half- and quarter-metals in rhombohedral trilayer graphene}.
\newblock \emph{\bibinfo{journal}{Nature}} \textbf{\bibinfo{volume}{598}}, \bibinfo{pages}{429--433} (\bibinfo{year}{2021}).
\newblock \urlprefix\url{https://www.nature.com/articles/s41586-021-03938-w}.
\newblock \bibinfo{note}{Number: 7881 Publisher: Nature Publishing Group}.

\bibitem{zhao_realization_2024}
\bibinfo{author}{Zhao, W.} \emph{et~al.}
\newblock \bibinfo{title}{Realization of the {Haldane} {Chern} insulator in a moiré lattice}.
\newblock \emph{\bibinfo{journal}{Nature Physics}} \textbf{\bibinfo{volume}{20}}, \bibinfo{pages}{275--280} (\bibinfo{year}{2024}).
\newblock \urlprefix\url{https://www.nature.com/articles/s41567-023-02284-0}.
\newblock \bibinfo{note}{Publisher: Nature Publishing Group}.

\bibitem{kometter_hofstadter_2023}
\bibinfo{author}{Kometter, C.~R.} \emph{et~al.}
\newblock \bibinfo{title}{Hofstadter states and re-entrant charge order in a semiconductor moiré lattice}.
\newblock \emph{\bibinfo{journal}{Nature Physics}} \bibinfo{pages}{1--7} (\bibinfo{year}{2023}).
\newblock \urlprefix\url{https://www.nature.com/articles/s41567-023-02195-0}.
\newblock \bibinfo{note}{Publisher: Nature Publishing Group}.

\bibitem{ghiotto_stoner_2024}
\bibinfo{author}{Ghiotto, A.} \emph{et~al.}
\newblock \bibinfo{title}{Stoner instabilities and {Ising} excitonic states in twisted transition metal dichalcogenides} (\bibinfo{year}{2024}).
\newblock \urlprefix\url{http://arxiv.org/abs/2405.17316}.
\newblock \bibinfo{note}{ArXiv:2405.17316 [cond-mat]}.

\bibitem{park_ferromagnetism_2024}
\bibinfo{author}{Park, H.} \emph{et~al.}
\newblock \bibinfo{title}{Ferromagnetism and {Topology} of the {Higher} {Flat} {Band} in a {Fractional} {Chern} {Insulator}} (\bibinfo{year}{2024}).
\newblock \urlprefix\url{http://arxiv.org/abs/2406.09591}.
\newblock \bibinfo{note}{ArXiv:2406.09591}.

\bibitem{streda_theory_1982}
\bibinfo{author}{Streda, P.}
\newblock \bibinfo{title}{Theory of quantised {Hall} conductivity in two dimensions}.
\newblock \emph{\bibinfo{journal}{Journal of Physics C: Solid State Physics}} \textbf{\bibinfo{volume}{15}}, \bibinfo{pages}{L717} (\bibinfo{year}{1982}).
\newblock \urlprefix\url{https://dx.doi.org/10.1088/0022-3719/15/22/005}.

\bibitem{eisenstein_negative_1992}
\bibinfo{author}{Eisenstein, J.~P.}, \bibinfo{author}{Pfeiffer, L.~N.} \& \bibinfo{author}{West, K.~W.}
\newblock \bibinfo{title}{Negative compressibility of interacting two-dimensional electron and quasiparticle gases}.
\newblock \emph{\bibinfo{journal}{Physical Review Letters}} \textbf{\bibinfo{volume}{68}}, \bibinfo{pages}{674--677} (\bibinfo{year}{1992}).
\newblock \urlprefix\url{https://link.aps.org/doi/10.1103/PhysRevLett.68.674}.

\bibitem{feldman_fractional_2013}
\bibinfo{author}{Feldman, B.~E.} \emph{et~al.}
\newblock \bibinfo{title}{Fractional {Quantum} {Hall} {Phase} {Transitions} and {Four}-{Flux} {States} in {Graphene}}.
\newblock \emph{\bibinfo{journal}{Physical Review Letters}} \textbf{\bibinfo{volume}{111}}, \bibinfo{pages}{076802} (\bibinfo{year}{2013}).
\newblock \urlprefix\url{https://link.aps.org/doi/10.1103/PhysRevLett.111.076802}.
\newblock \bibinfo{note}{Publisher: American Physical Society}.

\bibitem{zondiner_cascade_2020}
\bibinfo{author}{Zondiner, U.} \emph{et~al.}
\newblock \bibinfo{title}{Cascade of phase transitions and {Dirac} revivals in magic-angle graphene}.
\newblock \emph{\bibinfo{journal}{Nature}} \textbf{\bibinfo{volume}{582}}, \bibinfo{pages}{203--208} (\bibinfo{year}{2020}).
\newblock \urlprefix\url{https://www.nature.com/articles/s41586-020-2373-y}.

\bibitem{zhu_voltage-controlled_2020}
\bibinfo{author}{Zhu, J.}, \bibinfo{author}{Su, J.-J.} \& \bibinfo{author}{MacDonald, A.}
\newblock \bibinfo{title}{Voltage-{Controlled} {Magnetic} {Reversal} in {Orbital} {Chern} {Insulators}}.
\newblock \emph{\bibinfo{journal}{Physical Review Letters}} \textbf{\bibinfo{volume}{125}}, \bibinfo{pages}{227702} (\bibinfo{year}{2020}).
\newblock \urlprefix\url{https://link.aps.org/doi/10.1103/PhysRevLett.125.227702}.
\newblock \bibinfo{note}{Publisher: American Physical Society}.

\bibitem{redekop_direct_2024}
\bibinfo{author}{Redekop, E.} \emph{et~al.}
\newblock \bibinfo{title}{Direct magnetic imaging of fractional {Chern} insulators in twisted {MoTe}\$\_2\$ with a superconducting sensor} (\bibinfo{year}{2024}).
\newblock \urlprefix\url{http://arxiv.org/abs/2405.10269}.
\newblock \bibinfo{note}{ArXiv:2405.10269 [cond-mat]}.

\bibitem{knuppel_correlated_2024}
\bibinfo{author}{Knüppel, P.} \emph{et~al.}
\newblock \bibinfo{title}{Correlated states controlled by tunable van {Hove} singularity in moiré {WSe2}} (\bibinfo{year}{2024}).
\newblock \urlprefix\url{http://arxiv.org/abs/2406.03315}.
\newblock \bibinfo{note}{ArXiv:2406.03315}.

\bibitem{zhang_polarization-driven_2024}
\bibinfo{author}{Zhang, X.-W.} \emph{et~al.}
\newblock \bibinfo{title}{Polarization-driven band topology evolution in twisted {MoTe}\$\_2\$ and {WSe}\$\_2\$} (\bibinfo{year}{2024}).
\newblock \urlprefix\url{http://arxiv.org/abs/2311.12776}.
\newblock \bibinfo{note}{ArXiv:2311.12776 [cond-mat]}.

\bibitem{wang_valley-_2017}
\bibinfo{author}{Wang, Z.}, \bibinfo{author}{Shan, J.} \& \bibinfo{author}{Mak, K.~F.}
\newblock \bibinfo{title}{Valley- and spin-polarized {Landau} levels in monolayer {WSe2}}.
\newblock \emph{\bibinfo{journal}{Nature Nanotechnology}} \textbf{\bibinfo{volume}{12}}, \bibinfo{pages}{144--149} (\bibinfo{year}{2017}).
\newblock \urlprefix\url{https://www.nature.com/articles/nnano.2016.213}.
\newblock \bibinfo{note}{Number: 2 Publisher: Nature Publishing Group}.

\bibitem{shi_bilayer_2022}
\bibinfo{author}{Shi, Q.} \emph{et~al.}
\newblock \bibinfo{title}{Bilayer {WSe2} as a natural platform for interlayer exciton condensates in the strong coupling limit}.
\newblock \emph{\bibinfo{journal}{Nature Nanotechnology}} \textbf{\bibinfo{volume}{17}}, \bibinfo{pages}{577--582} (\bibinfo{year}{2022}).
\newblock \urlprefix\url{https://www.nature.com/articles/s41565-022-01104-5}.
\newblock \bibinfo{note}{Number: 6 Publisher: Nature Publishing Group}.

\bibitem{qiu_interaction-driven_2023}
\bibinfo{author}{Qiu, W.-X.}, \bibinfo{author}{Li, B.}, \bibinfo{author}{Luo, X.-J.} \& \bibinfo{author}{Wu, F.}
\newblock \bibinfo{title}{Interaction-driven topological phase diagram of twisted bilayer {MoTe}\$\_2\$} (\bibinfo{year}{2023}).
\newblock \urlprefix\url{http://arxiv.org/abs/2305.01006}.
\newblock \bibinfo{note}{ArXiv:2305.01006 [cond-mat]}.

\bibitem{foutty_tunable_2023}
\bibinfo{author}{Foutty, B.~A.} \emph{et~al.}
\newblock \bibinfo{title}{Tunable spin and valley excitations of correlated insulators in \${\textbackslash}{Gamma}\$-valley moiré bands}.
\newblock \emph{\bibinfo{journal}{Nature Materials}} \bibinfo{pages}{1--6} (\bibinfo{year}{2023}).
\newblock \urlprefix\url{https://www.nature.com/articles/s41563-023-01534-z}.
\newblock \bibinfo{note}{Publisher: Nature Publishing Group}.

\bibitem{mcgilly_visualization_2020}
\bibinfo{author}{McGilly, L.~J.} \emph{et~al.}
\newblock \bibinfo{title}{Visualization of moiré superlattices}.
\newblock \emph{\bibinfo{journal}{Nature Nanotechnology}} \textbf{\bibinfo{volume}{15}}, \bibinfo{pages}{580--584} (\bibinfo{year}{2020}).
\newblock \urlprefix\url{https://www.nature.com/articles/s41565-020-0708-3}.
\newblock \bibinfo{note}{Number: 7 Publisher: Nature Publishing Group}.

\end{thebibliography}



\section{Methods}
\subsection{Sample fabrication}
The tWSe$_2$ device was fabricated using standard dry transfer techniques. We use a poly(bisphenol A carbonate) (PC)/polydimethylsiloxane (PDMS) stamp to pick up a hexagonal boron nitride (hBN) flake (15 nm thick). We then sequentially pick up the first half of a monolayer WSe$_2$ flake (exfoliated from HQ Graphene source), using the hBN to tear the flake in two, followed by the second half after rotating the substrate by an angle of $1.5^\circ$. Separately, we prepare a stack with a bottom hBN (13 nm) atop a graphite (5 nm) back gate, upon which we deposit pre-patterned Cr/Pt contacts (2 nm / 8 nm) using standard electron-beam lithography techniques. This is annealed at $\approx 300\ ^\circ$C for 8 hours to clean polymer and resist residues both before and after pre-patterning of contacts. The tWSe$_2$ stack is then dropped onto the pre-patterned contacts. To improve electrical contact to the tWSe$_2$, we fabricate local ``contact-gates" over the Pt contacts, while leaving the rest of the sample ungated on top for access with the SET \cite{foutty_tunable_2023}.

\subsection{SET Measurements}

The SET sensor was fabricated by evaporating aluminum onto a pulled quartz rod, with an estimated diameter at the apex of 50-100 nm. The SET ``tip" is brought to about $50$ nm above the sample surface. Scanning SET measurements were performed in a Unisoku USM1300 scanning probe microscope with a customized microscope head. a.c.~excitations (2-5 mV peak-to-peak amplitude) were applied to both sample and back gate at distinct frequencies between 200 and 400 Hz. We then measure inverse compressibility $\textrm{d}\mu/\textrm{d}n \propto I_{\textrm{BG}} / I_{\textrm{2D}}$ where $ I_{\textrm{BG}}$ and $I_{\textrm{2D}}$ are measurements of the SET current demodulated at respective frequencies of the back gate and sample excitations \cite{yu_correlated_2022}. Except where otherwise noted (Fig.~\ref{fig:Fig5} and Extended Data Fig.~\ref{fig:DFieldDep_OtherTransitions}), a d.c.~offset voltage $V_{\textrm{2D}}$ is applied to the sample to maintain the working point of the SET at its maximum sensitivity point within a Coulomb blockade oscillation fringe chosen so that the tip does not gate the sample. This minimizes tip-induced doping and provides a direct measurement of $\mu(n)$ at d.c.~timescales. Depending on measurement location, we measure a small ($<1\times 10^{-11}$ meV cm$^{2}$) difference in d$\mu$/d$n$ between a.c.~and d.c.~measurements, which we subtract from the a.c.~in all data presented in the main text. The contact gates are held at a large, negative voltage throughout the measurement to maintain good electrical contact across a range of hole dopings. All SET measurements are taken at $T = 330$~mK unless otherwise noted.

\subsection{Determination of $M_z$}
In our experimental setup, we measure both d$\mu$/d$n$ (using an a.c.~technique) and $\mu$ (using a d.c.~technique) with carrier density varied as a fast axis and magnetic field as a slow axis. To extract d$\mu$/d$B$, we use the d.c.~measurement of $\mu$, which is given by $-1\times V_{\rm{2D}}$, where $V_{\rm{2D}}$ is the voltage applied to the sample to maintain constant current through the SET. This is preferable to numerically integrating the a.c.~measured d$\mu$/d$n$ because small offsets in the a.c.~signal lead to line-by-line differences in the integrated $\mu$. Because the d.c.~measurement suffers from higher 1/$f$ noise, we apply a smoothing Savitzky-Golay filter to mitigate these effects (Supplementary Fig.~S8). In order to compare data taken at distinct magnetic fields, we need to set a reference point (zero) of $\mu$ as we measure the chemical potential. We choose to set $\mu(\nu = -1) = 0$, such that the chemical potential is zero at the center of the $\nu = -1$ gap. By repeating this at many magnetic fields, we build up a function $\mu(n,B)$. For a more detailed discussion of this convention, see Supplementary Sec. 5.

In order to extract the magnetization $M_z$, we make use of the Maxwell relation relating the chemical potential $\mu$ to the magnetization $M_z$, namely
\begin{equation}
\left(\frac{\text{d}\mu}{\text{d}B}\right)_n = -\left(\frac{\text{d}M_z}{\text{d}n}\right)_B
\end{equation} where we specify that the magnetization that we probe is in the $z$-direction due to the direction of the magnetic field. Note that the choice of the zero of the chemical potential has the effect of setting d$M_z$/d$n$ to zero at $\nu = -1$. In order to extract the magnetization $M_z$ from the derivative d$M_z$/d$n$, we numerically integrate from the carrier density of $\nu = -1$, so all our measurements of magnetization are relative to the total magnetization at $\nu = -1$:
\begin{equation}
    M_z(\nu) = \int_{\nu = -1}^\nu \left(\frac{\text{d}M_z}{\text{d}n}\right) dn
\end{equation}

\subsection{Density and twist angle determination}

The target angle during stacking ($1.5^\circ$) sets the approximate twist angle of the device. However, relaxation during fabrication can alter the final device configuration. To ensure that the tWSe$_2$ does not relax substantially or is significantly strained after picking it up, we perform piezoelecric force microscoy (PFM) using a Bruker Icon atomic force microscope on the sample supported by the PC slide before setting the stack onto the prepatterned Pt contacts \cite{mcgilly_visualization_2020}. These measurements indicate minimal relaxation and limited strain \cite{foutty_mapping_2024}.

We use SET measurements to precisely determine the local twist angle. From the slopes of the Hofstadter states, we can calibrate the conversion between back gate voltage ${V}_{\rm{BG}}$ and carrier density $n$. We then use the integer gaps we measure (e.g.~at $\nu = -1,-2$, and $-3$) and/or the corresponding Hofstadter intercepts (in cases where there is no zero-field gap) to determine the density corresponding to filling one hole per moir\'e unit cell. From this density $n_{\rm{muc}}$, we convert to twist angle $\theta$ via $1/n_{\rm{muc}} = \frac{\sqrt{3}a_{\rm{WSe}_2}^2}{4-4\cos\theta}$ and subsequently moir\'e wavelength via $\lambda_m = \frac{a_{\rm{WSe}_2}}{2\sin(\theta/2)}$, where $a_{\rm{WSe}_2}=0.328$ nm.

\subsection{Electric field tuning with SET}
We follow the approach described in \cite{foutty_mapping_2024}. We model the system by treating the tip as a parallel plate capacitor: $D_{\textrm{eff}}= \frac{1}{2\epsilon_0}(C_t(0-(V_{\rm{2D}}-V_{\rm{{fb}}})) - C_b (V_{\rm BG}-V_{\rm 2D}-V_0))$, where $C_{b(t)}$ is the back (top) gate capacitance, $V_{\rm 2D}$ and $V_{\rm BG}$ are the d.c.~voltages applied to sample and back gate, $V_{\rm fb}$ is the ``flat-band" voltage at which the tip and sample are work-function-compensated, and $V_0$ is the voltage at which back gate and sample are work-function-compensated (equivalent to the voltage of the WSe$_{2}$ band edge). While $C_t$ depends on the height of the tip, we experimentally extract $C_t \approx 0.045C_b$ for the measurements shown based on the shifts of constant-density features in the $V_{\rm BG}-V_{\rm 2D}$ plane. The data in Fig.~\ref{fig:Fig5}b-d are taken by initially fixing $V_{\rm 2D}$ at a particular voltage and then sweeping $V_{\rm BG}$, feeding back on the value of $V_{\rm 2D}$ while data is taken to maintain fixed current through the SET.

\subsection{Calculation of Hofstadter thermodynamics}
To calculate the Hofstadter spectrum of tWSe$_2$ shown in Fig.~\ref{fig:Fig2}d-e, we diagonalize the continuum model of Ref.~\cite{devakul_magic_2021} in a Landau level basis following standard methods described in references such as \cite{bistritzer_moire_2011,kometter_hofstadter_2023,kolar_hofstadter_2024}. We define the magnetic flux quantum $\Phi_0=\frac{h}{e}$ and the magnetic flux per moiré unit cell $\Phi=BA_{\rm{uc}}$, where $B$ is the magnetic field strength and $A_{\rm{uc}}$ is the moiré unit cell area. We compute the spectrum at rational values of magnetic flux per unit cell $\Phi/\Phi_0 = p/q$ where $p$ and $q$ are relatively prime. We sample rational values of $p/q$ by looping $q$ from 1 to $q_{\text{max}}=40$, then looping from $p$ from $1$ to $q$. We truncate our basis at a maximum Landau level index $n_{\text{max}}= \lceil \frac{E_{\text{cut}}}{\hbar\omega_c(\Phi)} \rceil$ where $\omega_c = \frac{eB}{m}$ is a function of the flux. We set $E_{\text{cut}} = 30 \hbar\omega_c(\Phi=\Phi_0) \approx 221$ meV at $\theta=1.44^{\circ}$. We choose a magnetic unit cell with primitive vectors $\bm{a}_1$ and $q\bm{a}_2$ where $\bm{a}_i$ are primitive moiré lattice vectors. We then impose (magnetic) periodic boundary conditions with respect to the boundary vectors $L \bm{a}_1$ and $L\bm{a}_2$ where $L = q\lceil \sqrt{N_{\text{uc},\text{min}}}/q \rceil$, thereby constraining the allowed magnetic crystal momentum values. We set $N_{\text{uc},\text{min}}=3000$. Note than $L$ must be a multiple of $q$ to fit an integer number of magnetic unit cells.

To compute the data shown in Fig.~\ref{fig:Fig2}c (and Figs. S4-S5 in the Supplementary Information), we consider a Stoner model with total energy
\begin{equation}\label{eq:StonerModel}
    E(N_{\uparrow},N_{\downarrow})= -\sum_{\sigma=\uparrow,\downarrow} \sum_{i=1}^{N_{\sigma}} \varepsilon_{\sigma,i} -g\frac{\mu_B B}{2}\left(N_{\uparrow} - N_{\downarrow} \right)+ \frac{U}{N_{uc}}N_{\uparrow}N_{\downarrow}
\end{equation}
Here $\varepsilon_{i,\sigma}$ is the $i^{th}$ highest single-electron eigenvalue of spin $\sigma$ excluding Zeeman energy (see Supplementary Fig.~S2), $\mu_B = \frac{e\hbar}{2m_e}$ is the Bohr magneton, and $g$ is phenomenological Landé factor. $N_{\sigma}$ is the number of $\emph{holes}$ with spin $\sigma$. To incorporate nonzero temperature and disorder effects phenomenologically, we add a random shift $\delta \varepsilon_{\sigma,i}$ sampled from the Gaussian probability distribution $P(\delta\varepsilon) = \frac{1}{\sqrt{2\pi (k_B \gamma)^2}}e^{-\frac{\delta\varepsilon^2}{2(k_B \gamma)^2}}$ to each $\varepsilon_{\sigma,i}$. In Fig.~\ref{fig:Fig2}c, $\gamma=1$ K, $g=16$, $\theta=1.44^{\circ}$, and $U=2$ meV. Also, we include an interlayer potential bias of $\Delta=4$ meV to approximate the bias intrinsic to our single-gated device structure. For each value of $N = N_{\uparrow} + N_{\downarrow}$, we compute $E(N) \equiv \min[E(N_{\uparrow},N-N_{\uparrow})]$ by brute minimization (computing all possible values and keeping the minimum). The chemical potential is $\mu(N) = -[E(N+1)-E(N)]$.

We now describe how we compute derivatives of the chemical potential with respect to $n$, d$\mu$/d$n$, and $B$, $\text{d}\mu/\text{d}B|_n=-\text{d}M_z/\text{d}n|_B$. For different values of $B$, the system size varies due to the requirement of commensurability with the magnetic unit cell. Therefore, the calculation described above in the previous paragraph produces $\mu (\nu,B)$ on an irregular grid of $B$ and $\nu=-\frac{N}{N_{\rm{uc}}}$ values where $N_{\rm{uc}}$ is the number of moiré unit cells. We thus define a regular grid of $1000$ $\nu$ values labeled $\nu_i$ ranging from $\nu_{\min}=-5$ to $\nu_{\max}=0$ and linearly interpolate the known $\mu$ data onto this grid as a function of $\nu$ for each $B$. Finally, we approximate $\frac{\partial \mu }{\partial B}(\nu_i,B_j) \approx \frac{\mu(\nu_i,B_{j+1}) - \mu(\nu_i,B_{j-1})}{B_{j+1} - B_{j-1}} $ ($B_j$ is the $j^{th}$ smallest value of $B$). Also, we approximate $\frac{\partial \mu}{ \partial n}(\nu_i,B_j) \approx - A \frac{\mu(\nu_i+1,B_j) - \mu(\nu_i,B_j)}{\nu_{i+1}-\nu_{i}}$ where $A=N_{\rm{uc}}\frac{\sqrt{3}}{2}a_M^2$ is the total system area.

\section{Data availability}
The data that supports the findings of this study are available from the corresponding authors upon reasonable request.

\section{Code availability}
The codes that support the findings of this study are available from the corresponding authors upon reasonable request.

\section{Acknowledgments}
We acknowledge discussions with Vladimir Calvera, Zhaoyu Han, and Liang Fu for collaboration on related work. A.P.R. also thanks Junkai Dong for helpful discussions. Scanning SET measurements were supported by the Department of Energy, Office of Basic Energy Sciences, award number DE-SC0023109. Transport experiments were supported by NSF-DMR-2237050. B.E.F. acknowledges a Cottrell Scholar Award. A.P.R. was supported by AFOSR Award No. FA9550- 22-1-0432 and by grant NSF PHY-2309135 to the Kavli Institute for Theoretical Physics (KITP). The theoretical calculations benefited from computing resources provided by the MIT SuperCloud and Lincoln Laboratory Supercomputing Center. K.W. and T.T. acknowledge support from the JSPS KAKENHI (Grant Numbers 21H05233 and 23H02052), the CREST (JPMJCR24A5), JST and World Premier International Research Center Initiative (WPI), MEXT, Japan. T.D. acknowledges a startup fund from Stanford University. Part of this work was performed at the Stanford Nano Shared Facilities (SNSF), supported by the National Science Foundation under award ECCS-2026822.

\section{Author contribution}
B.A.F. conducted the experimental measurements and fabricated the samples with assistance from C.R.K. B.A.F. and B.E.F. designed the experiment. A.P.R. conducted theoretical calculations. K.W. and T.T. provided the hBN crystals. B.E.F. and T.D. supervised the project. All authors participated in analysis of the data and writing of the manuscript. 

\section{Competing interests}
The authors declare no competing interest. 

\newpage
\onecolumngrid
\section{Extended Data Figures}
\renewcommand{\figurename}{Extended Data Figure}
\setcounter{figure}{0}

 \begin{figure*}[h]
    \renewcommand{\thefigure}{\arabic{figure}}
    \centering
    \includegraphics[scale=1.0]{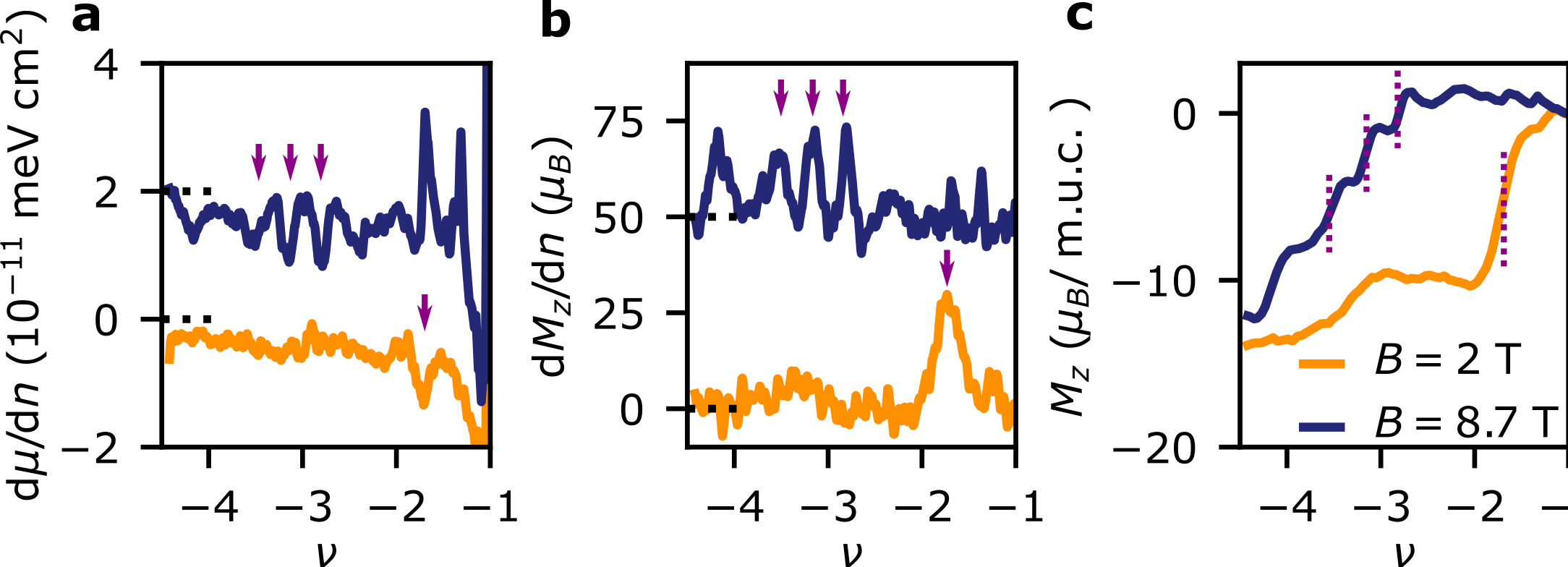}
    \caption{\textbf{Measurement of magnetization across phase transitions},  \textbf{a}-\textbf{b}, Linecuts of d$\mu$/d$n$ (\textbf{a}) and d$M_z$/d$n$ (\textbf{b}) at selected magnetic fields from the datasets presented in Fig.~\ref{fig:Fig1}d and Fig.~\ref{fig:Fig2}a. Purple arrows indicate positions of phase transitions. \textbf{c}, Magnetization $M_z$ relative to the magnetization at $\nu = -1$ as a function of $\nu$ at fixed $B$. This quantity is a numerical integration of \textbf{b}. The total change in magnetization in the two limits is similar because total number of carriers reversing spin is the same whether the underlying moir\'e bands have split into Hofstadter subbands or not.}
    \label{fig:MagLinetraces}
\end{figure*}

 \begin{figure*}[h]
    \renewcommand{\thefigure}{\arabic{figure}}
    \centering
    \includegraphics[scale=1.0]{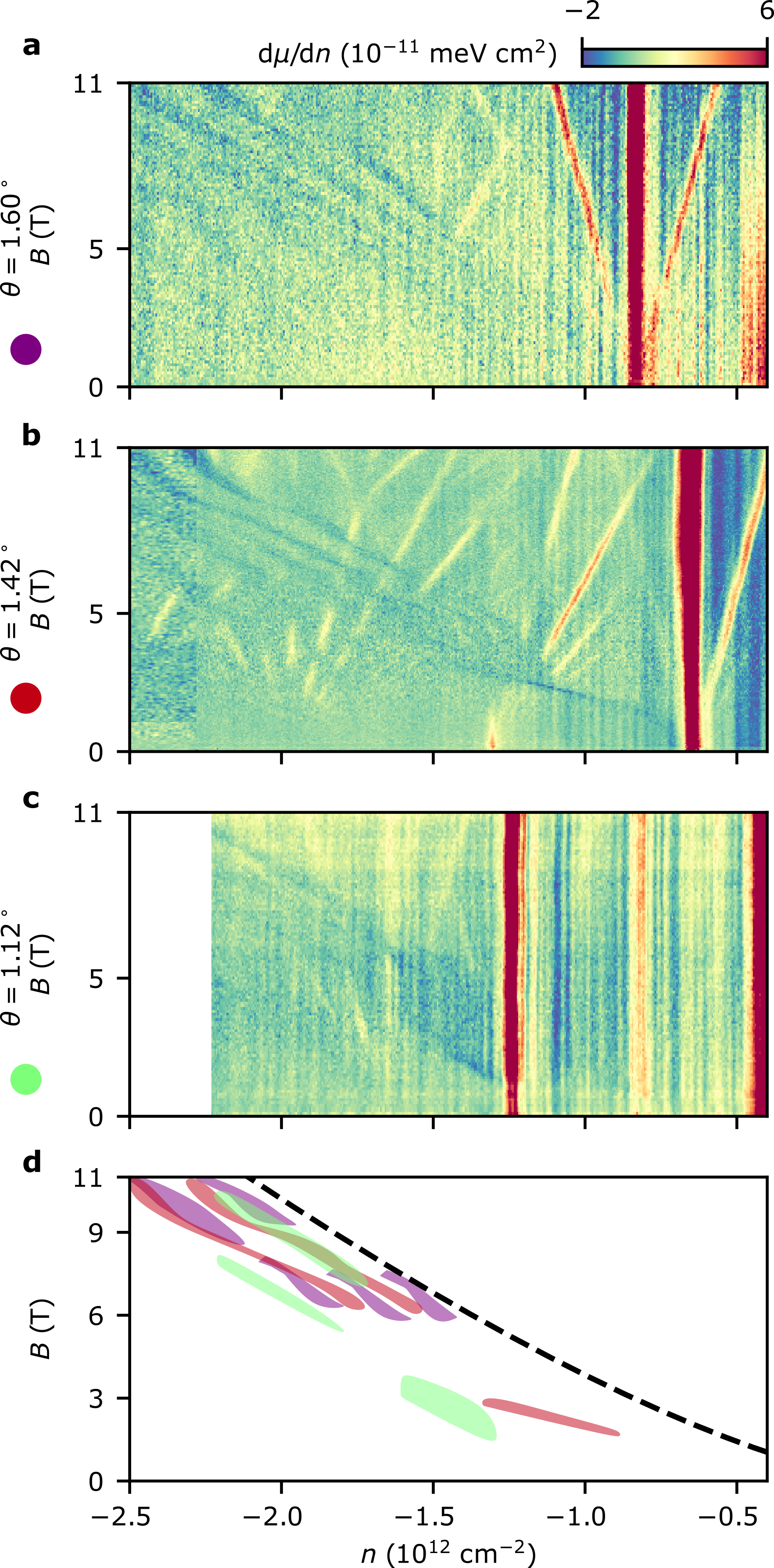}
    \caption{\textbf{Magnetic phase transitions at additional twist angles.} \textbf{a}-\textbf{c}, d$\mu$/d$n$ as a function of $n$ and $B$ at $\theta = 1.60^\circ$ (\textbf{a}), $\theta = 1.42^\circ$ (\textbf{b}) and $\theta = 1.12^\circ$ (\textbf{c}). \textbf{d}, Schematic showing the most prominent transitions from \textbf{a}-\textbf{c} alongside the monolayer WSe$_2$ transition line highlighted in Fig.~\ref{fig:Fig4}. }
    \label{fig:OtherTwistsdmudn}
\end{figure*}

 \begin{figure*}[h]
    \renewcommand{\thefigure}{\arabic{figure}}
    \centering
    \includegraphics[scale=1.0]{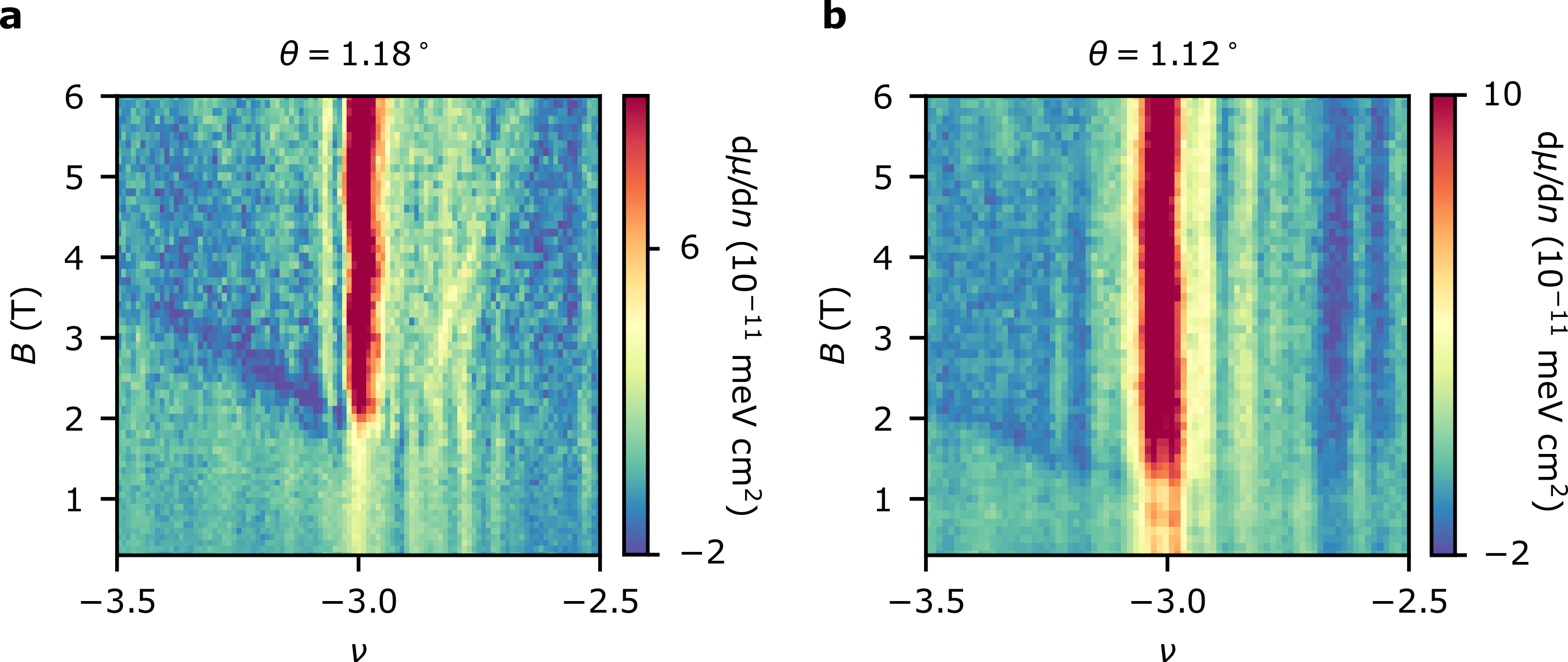}
    \caption{\textbf{Magnetic transitions near $\nu = -3$ at additional twist angles.} \textbf{a}-\textbf{b}, d$\mu$/d$n$ as a function of $\nu$ and $B$ around $\nu = -3$ at $\theta = 1.18^\circ$ (\textbf{a}) and $\theta = 1.12^\circ$ (\textbf{b}). At both twist angles, the $\nu = -3$ gap abruptly increases at a critical magnetic field, and there is an associated negative compressibility feature that extends to higher hole densities (at left).}
    \label{fig:Nu=-3_OtherTwists}
\end{figure*}

 \begin{figure*}[h]
    \renewcommand{\thefigure}{\arabic{figure}}
    \centering
    \includegraphics[scale=1.0]{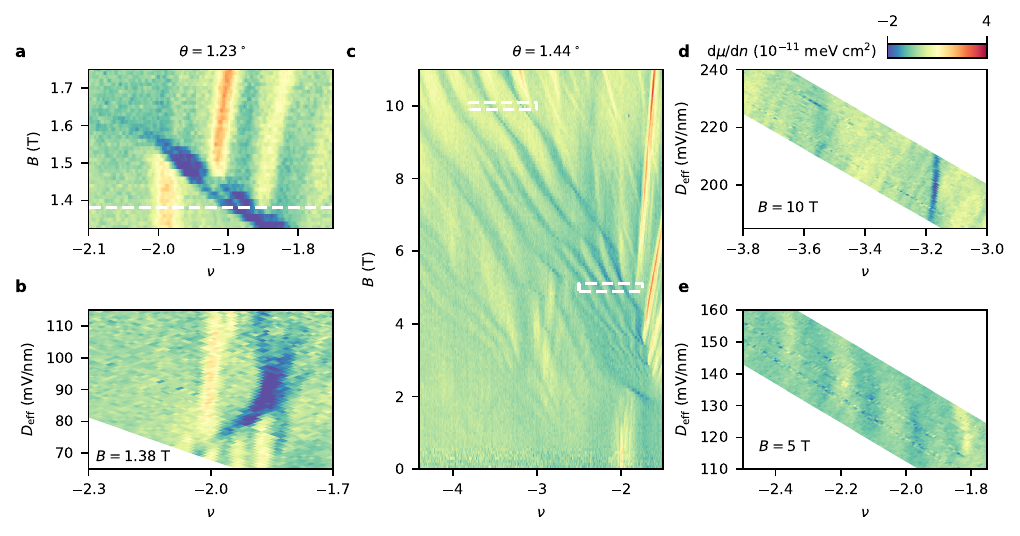}
    \caption{\textbf{Electric field effect at other magnetization transitions.} \textbf{a}, d$\mu$/d$n$ as a function of $\nu$ and $B$ around the magnetic transitions at $\nu = -2$ in a $\theta = 1.23^\circ$ location. \textbf{b}, d$\mu$/d$n$ as a function of $\nu$ and $D_{\rm{eff}}$ at $B = 1.38$ T (white dashed line in \textbf{a}). The magnetic transition is tuned by the changing displacement field. \textbf{c}, d$\mu$/d$n$ as a function of $\nu$ and $B$ in a $\theta = 1.44^\circ$ location, identical to Fig.~\ref{fig:Fig1} in the main text. \textbf{d}-\textbf{e}, d$\mu$/d$n$ at $B = 10$ T and $B = 5$ T, respectively, as a function of $\nu$ and $D_{\rm{eff}}$ around prominent magnetic transitions (dashed white boxes in \textbf{c}). There is no clear dispersion of the negative compressibility features with varying $D_{\rm{eff}}$, in contrast to the behavior in \textbf{b} and in Fig.~\ref{fig:Fig5} in the main text.}
    \label{fig:DFieldDep_OtherTransitions}
\end{figure*}


\end{document}


\title{Supplementary Information for:\\Magnetic Hofstadter cascade in a twisted semiconductor homobilayer}
\author{Benjamin A. Foutty}
\affiliation{Department of Physics, Stanford University, Stanford, CA 94305, USA}
\affiliation{Geballe Laboratory for Advanced Materials, Stanford, CA 94305, USA}

\author{Aidan P. Reddy}
\affiliation{Department of Physics, Massachusetts Institute of Technology, Cambridge, Massachusetts 02139, USA}

\author{Carlos R. Kometter}
\affiliation{Department of Physics, Stanford University, Stanford, CA 94305, USA}
\affiliation{Geballe Laboratory for Advanced Materials, Stanford, CA 94305, USA}

\author{Kenji Watanabe}
\affiliation{Research Center for Electronic and Optical Materials, National Institute for Materials Science, 1-1 Namiki, Tsukuba 305-0044, Japan}

\author{Takashi Taniguchi}
\affiliation{Research Center for Materials Nanoarchitectonics, National Institute for Materials Science, 1-1 Namiki, Tsukuba 305-0044, Japan}

\author{Trithep Devakul}
\affiliation{Department of Physics, Stanford University, Stanford, CA 94305, USA}

\author{Benjamin E. Feldman}
\email{bef@stanford.edu}
\affiliation{Department of Physics, Stanford University, Stanford, CA 94305, USA}
\affiliation{Geballe Laboratory for Advanced Materials, Stanford, CA 94305, USA}
\affiliation{Stanford Institute for Materials and Energy Sciences, SLAC National Accelerator Laboratory, Menlo Park, CA 94025, USA}
\maketitle
\tableofcontents

\section*{Supplementary Text}

\subsection{Toy model of Hofstadter subband crossings} \label{sec:PhenoModel}
Here we construct a toy model of the Hofstadter subband crossings which builds on the model in Ref.~\cite{kometter_hofstadter_2023}. Our goal with this approach is to develop and understand the minimal ingredients needed to capture the compressibility data in Fig.~1d in the main text.  We refer to the density of the spin-majority and spin-minority carriers as $n_\uparrow$ and $n_\downarrow$ respectively. We model the spin-majority carriers as a dispersive band with constant density of states. This is a simplification, as the true density of states will involve Hofstadter subbands, but it is justified because we are generally at higher fillings in this spin ($\nu_\uparrow > 2$) over the relevant range of parameters. In this range, gaps between moir\'e bands are reduced or absent (for example, we do not observe an incompressible state at $\nu=-4$ at zero magnetic field, indicating that the second and third moir\'e bands are overlapping). Likewise, the corresponding Hofstadter subbands from each moir\'e band are also overlapping, and in this limit, the density of states should approach twice that of monolayer WSe$_2$, which is approximately parabolic, with the factor of two coming from the coexistence of states from both layers \cite{wang_correlated_2020}. In contrast, the spin-minority states will have well resolved Hofstadter subbands, as we are filling states coming from the isolated first moir\'e band.

\begin{figure*}[h]
    \renewcommand{\thefigure}{S\arabic{figure}}
    \centering
    \includegraphics[scale=1.0]{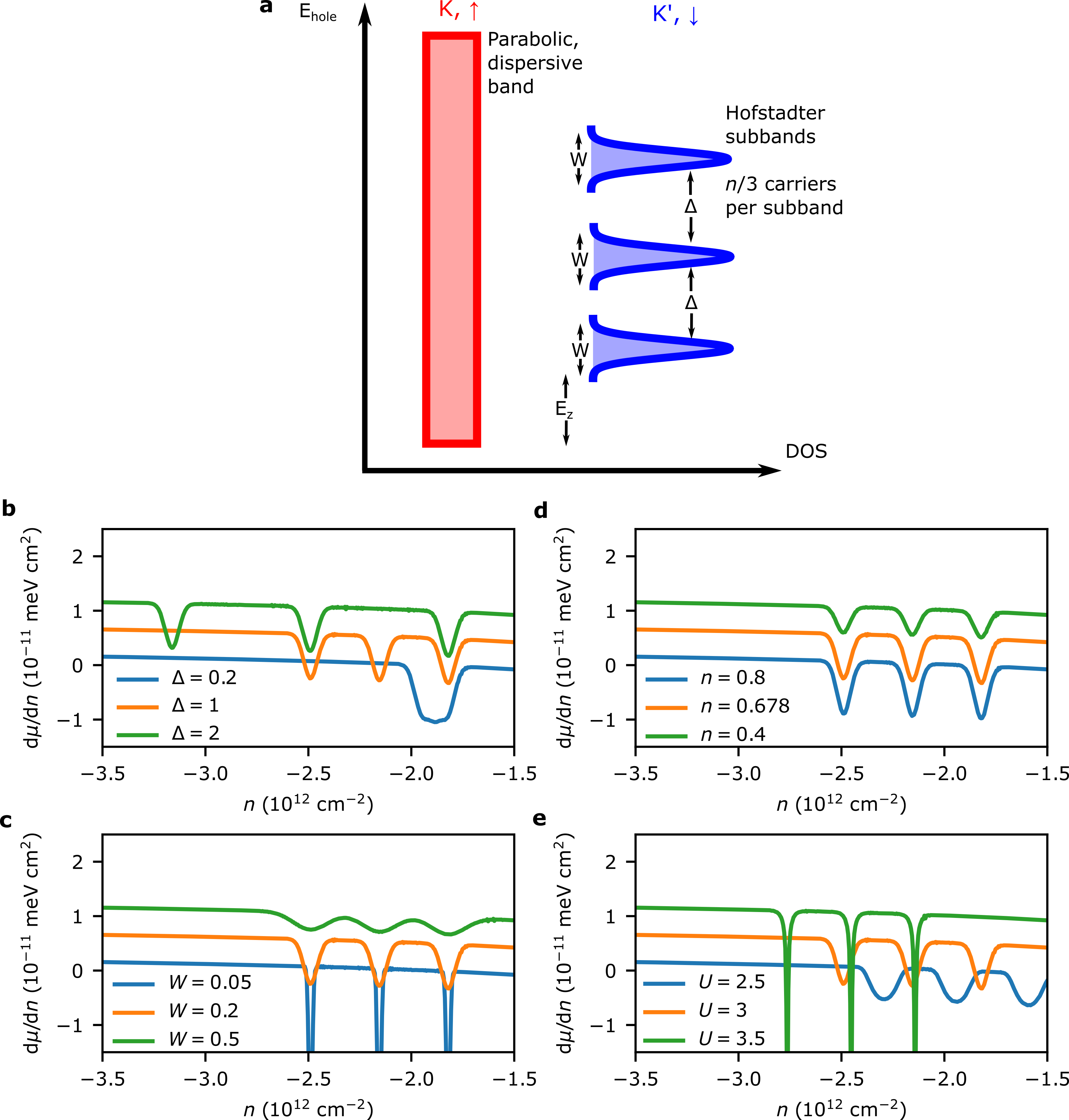}
    \caption{\textbf{Toy model of Hofstadter subband transitions.} \textbf{a}, Schematic of the model, which includes a parabolic, dispersive band in the spin-majority sector and a set of `Hofstadter subbands' in the spin-minority sector with a width $W$ and energy spacing $\Delta$ between them. There are assumed to be $n/3$ carriers per subband. \textbf{b}-\textbf{e}, We vary one of $\Delta$ (\textbf{b}), $W$ (\textbf{c}), $n$ (\textbf{d}), or $U$ (\textbf{e}) and plot the resulting d$\mu$/d$n$ from the model as discussed in the text. The default parameters for \textbf{b}-\textbf{e} are the following: $\Delta = 1$~meV, $W = 0.2$~meV, $n = 0.678\times 10^{12}$~cm$^{-2}$, and $U = 3 \times 10^{12}$~meV~cm$^{-2}$.}
    \label{fig:Stoner model}
\end{figure*}

At an arbitrary magnetic field, the structure of the spin-minority density of states will be complicated due the the rapid restructuring of Hofstadter subbands as a function of magnetic field. We therefore focus on rational flux $\Phi/\Phi_0 = \frac{1}{3}$ in which the first moir\'e band will split into three subbands of equal density. We take the density of states of each of these to be Gaussian-distributed with a width $W_i$, and spaced by gaps $\Delta_i$. We parameterize interactions via a Stoner $U$, as well as including a single-particle Zeeman energy, $E_Z = g \mu_B B$ where $g$ is the $g$-factor, $\mu_B$ is the Bohr magneton, and $B$ is the magnetic field (Fig.~\ref{fig:Stoner model}a). We also account for the slowly varying background negative compressibility in the experiment, which arises from exchange interactions and is unrelated to the cascade of Hofstadter transitions. The strength of these interactions are set by the dielectric environment and fundamental constants, which we incorporate in a single parameter $A_{ex}$ \cite{eisenstein_compressibility_1994,foutty_tunable_2023}.

Combining these ingredients, we write down the free energy of the system:
\begin{equation}
    E(n_\uparrow,n_\downarrow,B) = \frac{1}{2D_\uparrow} n_\uparrow^2 + E_{\rm{KE}}(n_\downarrow) + g\mu_B B |n_\downarrow| + U|n_\uparrow||n_\downarrow| - A_{ex}[(|n_\uparrow|)^{3/2}+(|n_\downarrow|)^{3/2}]
\end{equation}
$E_{\rm{KE}}$ describes the kinetic energy due to the density of states of the spin-minority subbands. We minimize this free energy with the constraint $n_\uparrow + n_\downarrow = n_{\rm{tot}}$ for fixed $n_{\rm{tot}}$. This yields a function $E(n_{\rm{tot}},B)$, and taking derivatives with respect to $n_{\rm{tot}}$ then gives the chemical potential $\mu$ and inverse electronic compressibility d$\mu$/d$n$ which can be compared to experiment. 

In Fig.~\ref{fig:Stoner model} we show the results from this model as we systematically change the parameters of the spin-minority density of states and the interaction strength to determine how each affects the strength and positions of the dips in d$\mu$/d$n$. In Fig.~\ref{fig:Stoner model}b, we vary the energy spacing between the spin-minority subbands. When the energy gaps between the subbands gets larger, the spacing of the negative d$\mu$/d$n$ gets larger. This is because there are more spin-majority states to fill within spin-minority subband gaps. When the spacings between the subbands shrink, closing the gaps, the negative d$\mu$/d$n$ feature merges into a single, broader dip. This naturally leads to the low-field behavior noted in the main text. In Fig.~\ref{fig:Stoner model}c, we vary the energy width of the subbands. As the spectral width of the subbands becomes narrower, the corresponding dips in d$\mu$/d$n$ also become much sharper. When the subband width is comparable to the energy spacing (shrinking the gaps) the dips are much broader. 

In Fig.~\ref{fig:Stoner model}d, we show the resulting d$\mu$/d$n$ as we change the total density across the three (evenly sized) subbands. This provides intuition for how the dips evolve when the number of subbands changes at a different magnetic flux, which causes a change in density per subband. The leading effect is to change the depth / size of the dips in d$\mu$/d$n$. In Fig.~\ref{fig:Stoner model}e, we change the Stoner parameter $U$. This has a sizeable effect of the sharpness of the dips in d$\mu$/d$n$, as this interaction term is really what determines how quickly spin-minority subbands will cross the Fermi level. Changing the Stoner parameter also strongly affects the density at which the first (lowest hole density) phase transition is observed. This is sensible because a larger Stoner interaction will increase the energy cost to begin doping into the spin-minority subbands, increasing the `effective' Zeeman energy \cite{foutty_anomalous_2024}. The onset of the phase transitions depends sensitively on the strength of the Zeeman energy and spin-majority density of states as well as the interaction strength. However, this simulation indicates that interaction effects will have a large impact on the exact positions of the phase transitions across the phase diagram, even if some of their other features (strength, relative spacing) can be reasonably well-understood through the single-particle spectrum itself.

In Section~\ref{HofstThermoSection}, we refine this phenomenological model by including the $B$-dependent density of states from Hofstadter calculations.

\subsection{Physical picture for measured negative $\rm{d}\mu/\rm{d}n$}
Here we expand on the physical picture behind the negative compressibility that is realized in experiment. Generally, negative compressibility of a thermodynamic system is incompatible with thermodynamic stability. Experimentally, we measure the 
component of the electronic compressibility coming from the internal energy of the 
material system under study (tWSe$_2$). Crucially, this is different from the electronic compressibility of the full sample-gate thermodynamic system, which additionally includes contributions from the gate-sample geometric capacitive energy and the internal energy of the gate. Therefore, a negative measured compressibility does not imply negative compressibility or instability of the full gate-sample system. 

There are two principle ways in which the measured compressibility can be negative. One mechanism is via a true first-order phase transition, i.e.~the existence of phase separation combined with finite size-effects or long-range interactions that prevent macroscopic domain formation \cite{zondiner_cascade_2020,yu_correlated_2022}. Alternatively, in numerous cases 
electron-electron interactions can lead to negative measured compressibility. One example of this is a two-dimensional electron gas at sufficiently low densities, in which exchange interactions lead to a negative d$\mu$/d$n$; this effect is what leads to the broad, negative background towards low hole densities in our data \cite{eisenstein_negative_1992,eisenstein_compressibility_1994,foutty_tunable_2023}.In the toy model described above (for sufficiently small $U$), interactions can cause spin-majority states to deplete while spin-minority states fill. This reduces the total spin-majority kinetic energy as $n$ increases, causing a negative d$\mu$/d$n$. Alternatively, a sufficiently large $U$ can cause a first order transition in which the magnetization and chemical potential change discontinuously. 

In our experiment, given that through most of the parameter space we measure broad and relatively shallow negative d$\mu$/d$n$ dips, we posit that most of these features are attributable to continuous transitions via (relatively weak) Stoner magnetism. However, at particular locations in the $\nu$-$B$ plane where d$\mu$/d$n$ becomes much sharper and more negative, interactions may be strong enough to drive first-order phase transitions and phase separation. We did not observe hysteresis (e.g., differences in sweeping the gate back and forth) in any measurements.

\subsection{Hofstadter thermodynamics}\label{HofstThermoSection}

\subsubsection{Hofstadter spectrum}

\begin{figure*}[h]
    \centering
    \renewcommand{\thefigure}{S\arabic{figure}}
    \includegraphics[width=0.6\linewidth]{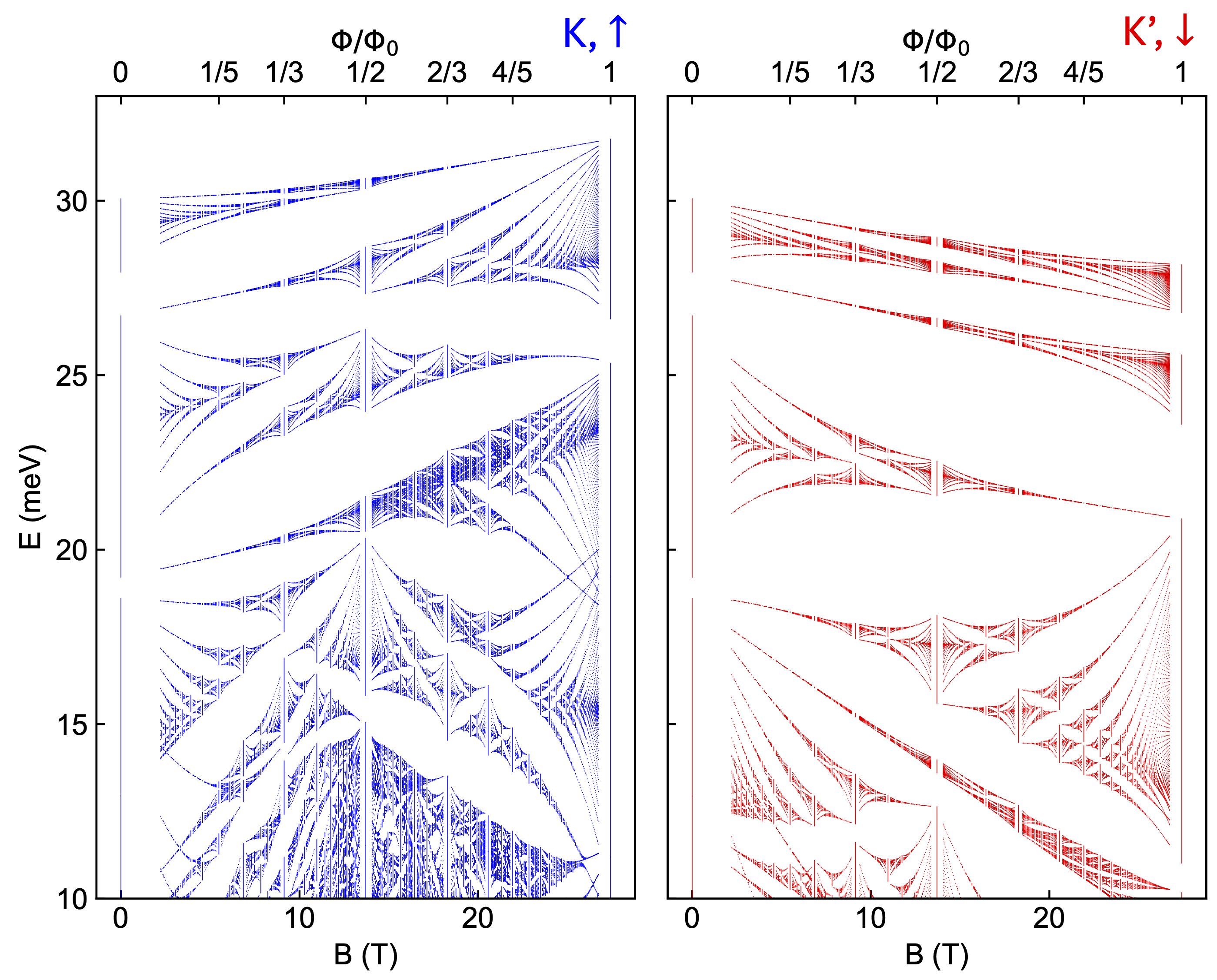}
    \caption{\textbf{Spin-resolved Hofstadter spectra.} Hofstadter spectra of spin-$\uparrow$ (left) and spin-$\downarrow$ (right) states plotted in the electron picture so that the valence band edge is at the top of the energy window shown. These are identical to the data in Fig.~2c in the main text, except that they exclude the Zeeman energy and cover a broader range of magnetic flux quanta per unit cell $\Phi/\Phi_0$ and densities. The twist angle is taken as $\theta=1.44^{\circ}$ and we introduce an interlayer potential bias of $\Delta=4$~meV.}
    \label{fig:HofstSpec}
\end{figure*}

In Fig.~\ref{fig:HofstSpec}, we separately plot the Hofstadter spectrum of each spin in tWSe$_2$ that we use in thermodynamic calculations. In comparison to Fig.~2d-e of the main text, Fig.~\ref{fig:HofstSpec} excludes Zeeman energy and shows wider range of energies and magnetic fields.

\begin{figure*}[h]
    \centering
    \renewcommand{\thefigure}{S\arabic{figure}}
    \includegraphics[width=0.6\linewidth]{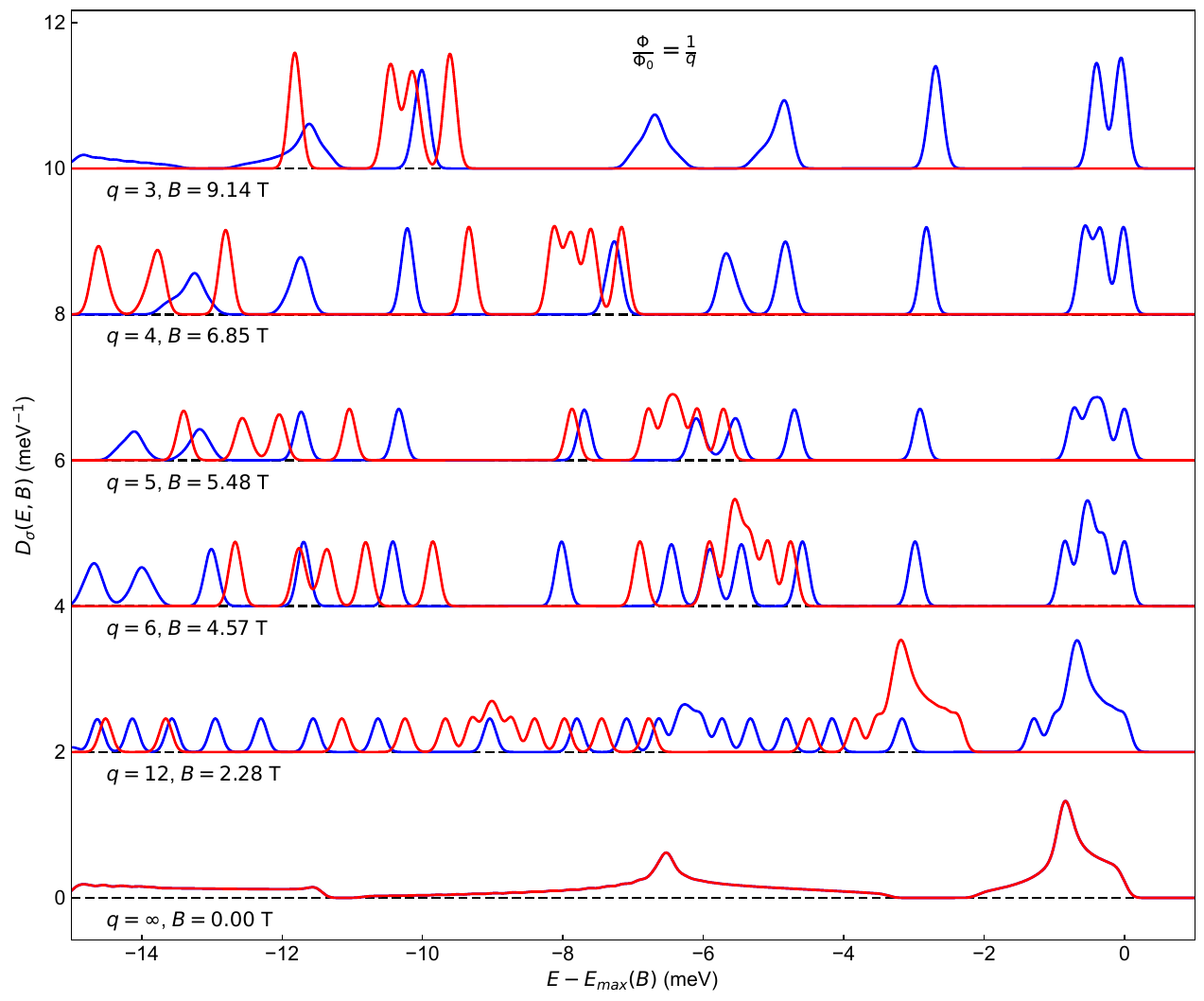}
    \caption{\textbf{Non-interacting density of states.} Non-interacting, spin-resolved density of states (including Zeeman energy) $D_{\sigma}(E,B) = \frac{1}{N_{\rm{uc}}}\sum_{i} \delta(E-\varepsilon_{\sigma,i}-\sigma g \mu_{B}B)$ at several values of $\Phi/\Phi_0=1/q$ with integer $q$. Here, 
    $\sigma = \pm 1/2$ for spin-majority (minority) states.
    We replace the $\delta$ function with a Gaussian of standard deviation $0.086$~meV (corresponding to a temperature of 1 K). $E_{\max}(B) = \varepsilon_{\uparrow,1} +  \frac{g}{2}\mu_{B}B$ is energy of the valence band edge. Note that the WSe$_2$ band gap is on the right side of this plot and that the bands are degenerate at $B=0$.}
    \label{fig:DOS}
\end{figure*}

In Fig.~\ref{fig:DOS}, we plot the spin-resolved, non-interacting density of states (DOS), $D_{\sigma}(E,B)$, at several values of $\Phi/\Phi_0=1/q$. Here, $\sigma = \pm \frac{1}{2}$ is an index denoting spin-majority (minority) states. At very low flux ($q=12$), Hofstadter gaps are washed out due to the finite broadening and the narrowness of the first moiré band at $B=0$. At larger flux such as $q=3$, sizable gaps emerge between Hofstadter subbands.

We note that several features of the Hofstadter spectra find simple explanations within the adiabatic approximation for the electronic structure of twisted $K$-valley TMD homobilayers \cite{morales-duran_magic_2024,kolar_hofstadter_2024}. In this approximation, the moiré superlattice at $B=0$ creates an emergent magnetic flux of one flux quantum per unit cell with opposite signs in opposite valleys. Consequently, applying an external magnetic field increases the net effective flux in one valley (here, $K$) and decreases it in the other ($K'$). 



\subsubsection{Simplified Stoner model}

While the toy model (Section \ref{sec:PhenoModel}) captures a significant portion of the experimental observations, here we introduce Stoner models of increasing complexity that incorporate realistic band structures. First in our `simplified' model, we include the $B$-dependent spin-$\downarrow$ DOS from the calculated Hofstadter spectrum (Fig. \ref{fig:HofstSpec}) with a constant spin-$\uparrow$ DOS. Specifically, we modify the Stoner model introduced in the Methods section of the main text by replacing the spin-$\uparrow$ energy levels $\varepsilon_{\uparrow,i}$ by uniformly sampling the constant DOS, $D_{\uparrow}(E,B) = A_{\rm{uc}} \frac{2\pi\hbar^2}{m^*}\Theta\left(\varepsilon_{\uparrow,1}(B) + g \frac{\mu_{B}}{2}B - E \right)$. This DOS corresponds to a parabolic valence band with the same $B$-dependent band edge as the full spin-$\uparrow$ spectrum and a twofold degeneracy coming from the layer degree of freedom.

\begin{figure*}[h]
    \renewcommand{\thefigure}{S\arabic{figure}}
    \centering
    \includegraphics[width=\linewidth]{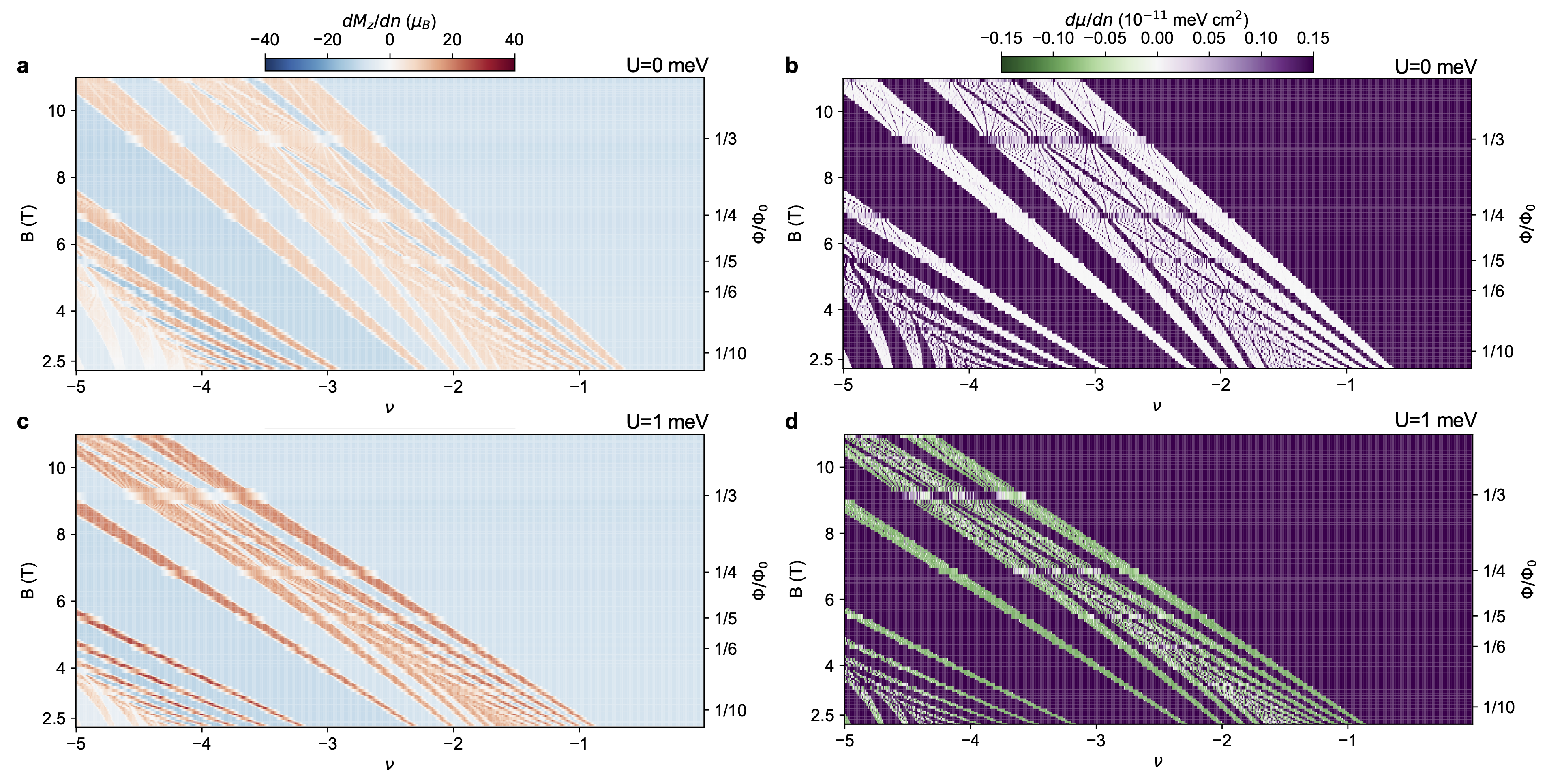}
    \caption{\textbf{Simplified Stoner model thermodynamics.} \textbf{a}-\textbf{b}, d$\mu$/d$n$ (\textbf{a}) and d$M_{z}$/d$n$ (\textbf{b}) calculated within the simplified Stoner model assuming a uniform spin-majority DOS but the full spin-minority Hofstadter spectrum, with interaction parameter $U=0$. \textbf{c}-\textbf{d}, Analogous calculations for $U=2$~meV. Here we include no broadening ($\gamma = 0$~K).}
    \label{fig:HofstThermo_Toy}
\end{figure*}

In Fig.~\ref{fig:HofstThermo_Toy}, we plot d$\mu/$d$n$ and d$M_{z}/$d$n$ of this simplified model in the $\nu$-$B$ plane. First, we consider the non-interacting case with $U=0$ in Fig.~\ref{fig:HofstThermo_Toy}a,b. The $\nu$-$B$ plane can be roughly divided into regions of two classes: blue regions where d$M_{z}$/d$n<0$ and d$\mu$/d$n$ is relatively large and red regions where d$M_{z}$/d$n>0$ and d$\mu$/d$n$ is relatively small. Blue regions are associated with spin-$\downarrow$ Hofstadter gaps and red regions are associated with spin-$\downarrow$ Hofstadter subbands. Red regions have lower d$\mu$/d$n$ because they have a higher DOS and, in the absence of interactions, d$\mu$/d$n(\mu,B) = A_{uc} \left[\sum_{\sigma} D_{\sigma}(\mu,B)\right]^{-1}$. The sign of d$M_{z}$/d$n$ is opposite in the two regions because single-particle states of opposite spin carry opposite magnetization (including both orbital and spin contributions). Figure~\ref{fig:HofstThermo_Toy}c,d includes a finite Stoner interaction $U=2$ meV. The main effects of this interaction are the same as those encountered in Section~\ref{sec:PhenoModel}: to drive higher spin polarization, to shrink the red regions where minority-spin Hofstadter subbands are filled, and to create negative compressibility in these red regions. We emphasize that the the Stoner model, Eq. (3) of the main text, is to be understood as a model for the internal energy of the tWSe$_2$ system.

\subsubsection{Full Stoner model}

\begin{figure*}[h]
    \renewcommand{\thefigure}{S\arabic{figure}}
    \centering
    \includegraphics[width=\linewidth]{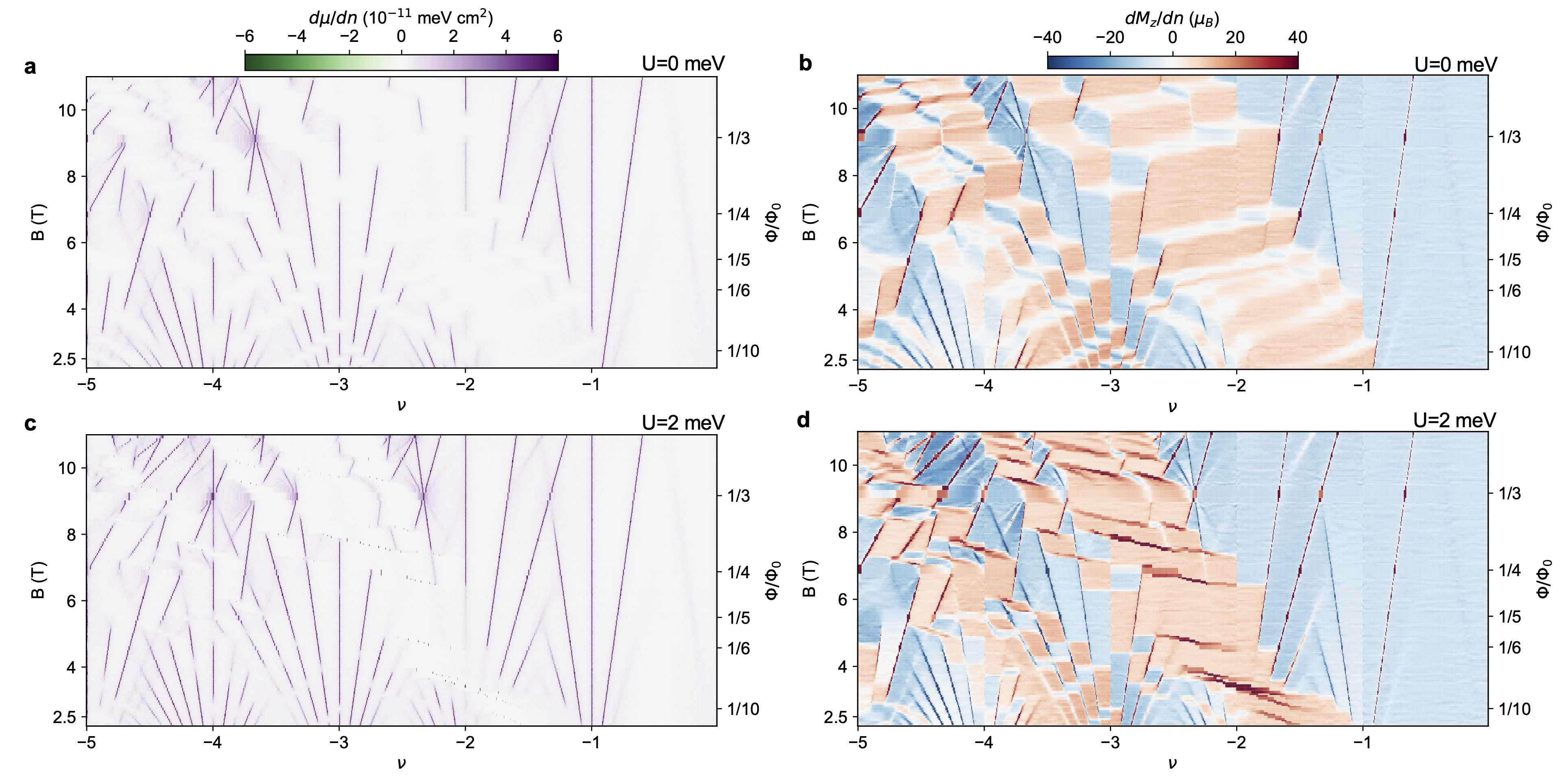}
    \caption{\textbf{Full Stoner model thermodynamics.} \textbf{a}-\textbf{b}, d$\mu$/d$n$ (\textbf{a}) and d$M_{z}$/d$n$ (\textbf{b}) of the Stoner model, Eq. 3 in the main text, with interaction parameter $U=0$. \textbf{c}-\textbf{d}, Analogous calculations for $U=2$~meV. We use a broadening parameter $\gamma = 1$~K. The separation of negative compressibility peaks in \textbf{c} is an artifact of the discreteness of the data as a function of $B$. These isolated features should form continuous lines.}
    \label{fig:HofstThermo_Full}
\end{figure*}

We now turn to the `full' Stoner model, including the spin-$\uparrow$ energy levels from the Hofstadter calculation in addition to the spin-$\downarrow$ states. Figure~\ref{fig:HofstThermo_Full}a,b shows the resulting d$\mu$/d$n$ and d$M_{z}$/d$n$ without interactions. In addition to the features present in the simplified model, Hofstadter gaps appear as sharp peaks in d$\mu$/d$n$ and $|$d$M_{z}$/d$n|$. Consistent with experiment, prominent gaps do not appear for $\nu>-1$, except for $(t,s)=(+1,-1)$ (where the location of the gap obeys $\nu=t\frac{\Phi}{\Phi_0}+s$). This is because the lowest moiré band at $B=0$ is narrow and becomes narrower as $B$ increases for spin-$\uparrow$, causing its small subgaps to disappear altogether in the presence of broadening. Next, the shape of the red regions (minority subbands) gains additional structure from the spin-majority DOS. Compared to the simplified model, some blue regions (minority subband gaps) intervening the red regions are shrunken or eliminated. In some cases this is this is because the number of spin-$\uparrow$ states in the spin-$\downarrow$ Hofstadter gaps is small and in other cases, [e.g.~around $(\nu,B)=(-2.5,6\text{ T})]$, it is because the spin-$\downarrow$ subband gaps close in the presence of broadening.

Figure~\ref{fig:HofstThermo_Full}c,d includes a finite Stoner interaction $U=2$ meV. Relative to the simplified model, the interaction leads to more pronounced peaks in d$M_{z}$/d$n$ associated with spin-majority depopulation. The most prominent such feature occurs near $(\nu,B)=(-2,3\text{ T})$. The experimental data in Fig.~2a of the main text shows a very similar feature at a similar position in the $(\nu,B)$ plane. Additionally, extended regions of negative compressibility in the simplified model are replaced by sharp lines of negative compressibility and positive d$M_{z}$/d$n$ (green, Fig.~\ref{fig:HofstThermo_Full}c). These sharp lines are associated with first order phase transitions in which the spin population changes discontinuously as a function of $\nu$ and occur due to the combination of strong DOS peaks and interactions. They contrast the broad negative compressibility features observed in the experimental data and in the simplified model with a uniform spin-majority DOS.

\subsubsection{Additional discussion}

Here we expand on the comparison between theoretical calculations and the experimental data. Our calculations exhibit more Hofstadter gaps than are apparent in the experimental data, particularly at larger hole density (Fig.~\ref{fig:HofstThermo_Full}). The calculations also show more first-order transitions, appearing as sharp lines of negative compressibility and positive d$M_{z}$/d$n$ in Fig.~\ref{fig:HofstThermo_Full}. In the experimental data, the regions of negative d$\mu$/d$n$ and positive d$M_z$/d$n$ are generally more separated than in the calculations. The simplified Stoner model presented above captures certain aspects of the experimental data, including the number and continuous character of the phase transitions, more faithfully than the full Stoner model. This suggests that the spin-$\uparrow$ DOS has weaker gaps than in the continuum model. In addition to extrinsic reasons such as disorder, intrinsic screening of the moir\'e potential, which should grow with $|\nu|$ and is not captured by the Stoner model, may be responsible for this difference. A broad DOS and the absence of large gaps at higher $|\nu|$ is consistent with the fact that in experiments at this twist angle, we do not observe an insulating state at $\nu = -4$ and $B = 0$ T, which would be a band insulator between the second and third moir\'e bands. At the same time, the full Stoner model more faithfully captures other aspects of the experimental data than the simplified uniform-DOS model, including the Hofstadter gaps and the strong, connected d$M_{z}$/d$n$ peak near $(\nu,B)\approx (-2, 3\text{ T})$ associated with rapid population of the first spin-minority moiré band.

One notable feature in the experimental data is a sizable phase transition  (around $B = 8$ T at $\nu = -4$ in Fig.~1c-d in the main text) just beyond the set of transitions highlighted in Fig.~3 in the main text. This transition occurs directly between the set of transitions attributed to the Hofstadter spectrum of the first moir\'e band and the weaker set of transitions at higher densities, which comes from higher moir\'e bands. This is consistent with an `extra' Landau level (LL) pinned to the bottom of the moir\'e band of the spin-minority Hofstadter spectrum (Fig.~\ref{fig:HofstSpec}) that stems from the underlying topology of the band structure. This `extra' LL is also apparent in the thermodynamic calculations shown Figs. \ref{fig:HofstThermo_Toy},\ref{fig:HofstThermo_Full}. That it is well-separated from the rest of the states tracing back to the first moir\'e band crossing is consistent with the fact that the underlying ($B = 0$) band structure is near a topological phase transition, as discussed in detail in Ref.~\cite{foutty_mapping_2024}.

\subsection{Transport measurements in a second device with $\theta = 2.2^\circ$}
\begin{figure*}[h]
    \renewcommand{\thefigure}{S\arabic{figure}}
    \centering
    \includegraphics[scale=1.0]{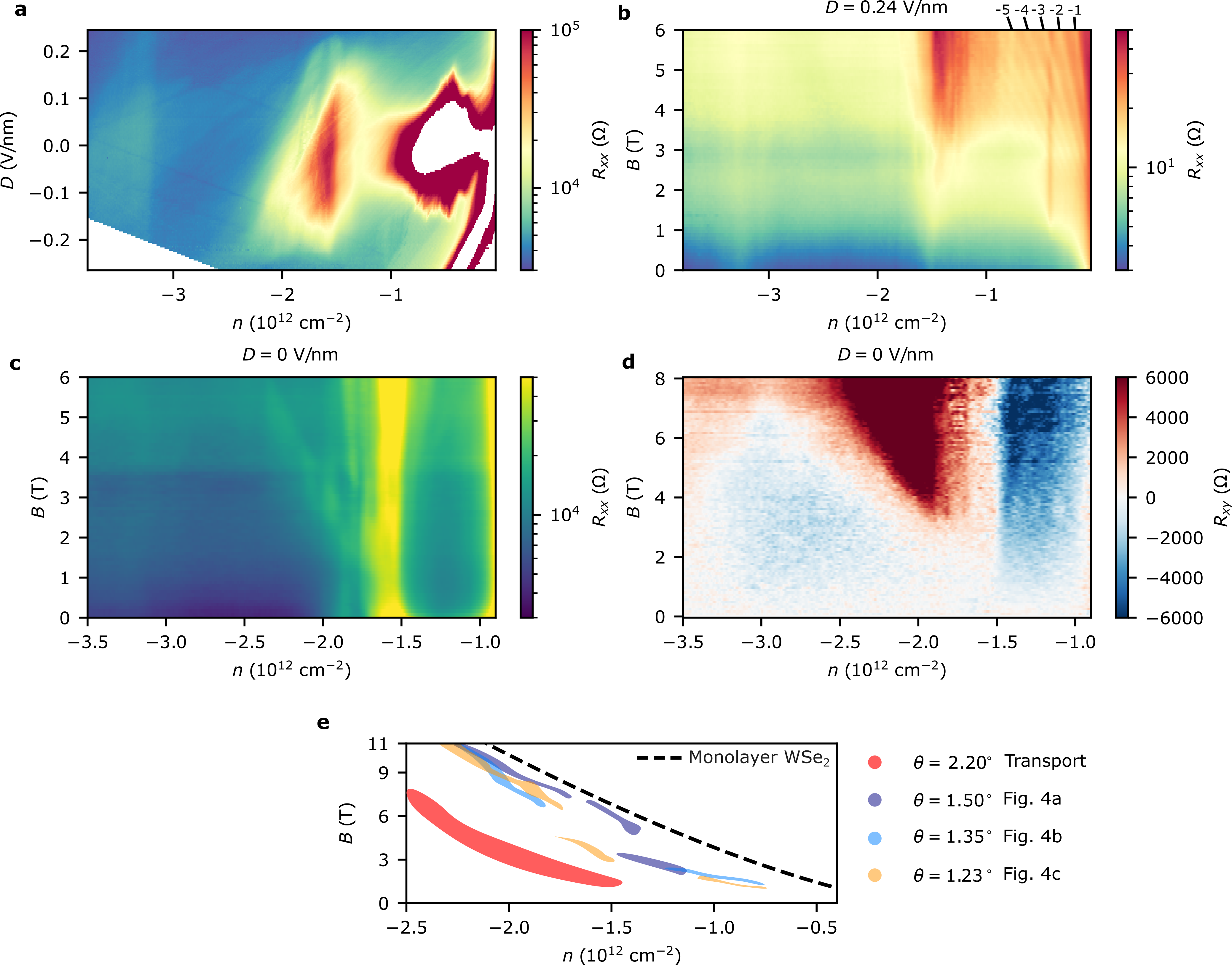}
    \caption{\textbf{Magnetotransport data from a sample with $\theta = 2.2^\circ$.} \textbf{a}, Longitudinal resistance $R_{xx}$ as a function of $n$ and $D$ at $B = 0$ T and $T = 350$ mK. \textbf{b}, $R_{xx}$ as a function of $n$ and $B$ at $D  =0.24$ V/nm. At a large displacement field, the insulating gap at $\nu = -1$ is weakened, and Landau levels tracing back to the band edge become more prominent. \textbf{c}-\textbf{d}, $R_{xx}$ (\textbf{c}) and $R_{xy}$ (\textbf{d}) as a function of $B$ at $D = 0$ V/nm. Data is symmetrized and anti-symmetrized using data from positive and negative magnetic field. \textbf{e}, Schematic showing the onset of magnetic phase transitions (at lowest hole density) for each twist angle from Fig.~4a-c, the transition extracted from transport, and the approximate location of the transition between full and partial spin polarization in monolayer WSe$_2$ for comparison.}
    \label{fig:Transport}
\end{figure*}

To explore over what range of twist angles the onset of magnetic transitions is approximately constant, we performed magnetotransport measurements in a separate device, which has a higher twist angle of $\theta = 2.2^\circ$. The sample is dual-gated with top and bottom graphite gates, with electrical contacts made via Pt contacts with contact gates to locally dope the contact regions \cite{park_observation_2023}. Due to the global nature of the transport measurement, we inherently average over the twist angle disorder in the sample. However, we find that this sample has sufficiently low inhomogeneity to resolve clear features in transport. In Fig.~\ref{fig:Transport}a, we show the longitudinal resistance $R_{xx}$ at $B = 0$ and $T = 350$ mK as a function of $n$ and out-of-plane displacement field $D$. We observe a prominent resistive feature at $n = -1.55 \times 10^{12}$ cm$^{-2}$, and a weaker resistive feature at $n = -3.1\times 10^{12}$ cm$^{-2}$. We associate these with insulating states at $\nu = -1$ and $\nu = -2$, consistent with prior reports of transport in this system \cite{wang_correlated_2020,kang_observation_2024,xia_unconventional_2024,knuppel_correlated_2024}. We determine the band edge ($n = 0$) filling and gate capacitances from singly-degenerate quantum oscillations measured at high displacement field and high magnetic field (Fig.~\ref{fig:Transport}b). In Fig.~\ref{fig:Transport}c-d, we show $R_{xx}$ and $R_{xy}$ measured in an out-of-plane magnetic field. The resistivity of the insulating state at $\nu = -1$ first weakens and then grows with increasing magnetic field, and there is a sign change of the Hall resistance crossing $\nu = -1$. There is also a broad resistive feature beginning around $\nu = -1$ at $B = 2$ T and sweeping up to higher hole densities at higher magnetic fields, which is concurrent with a sharp change in sign of $R_{xy}$. We interpret this as a moir\'e band crossing similar to the results in the main text. We also note that the observations in transport are similar to recent reports in tMoTe$_2$ \cite{park_ferromagnetism_2024}. In Fig.~\ref{fig:Transport}e, we show the position of this feature in $n$-$B$ space in comparison to those discussed in the main text. While the overall shape is similar, the transition at $\theta = 2.2^\circ$ occurs at relatively higher hole density. So even though we observe essentially no dependence on twist angle between $\theta = 1.1^\circ$ and $\theta = 1.6^\circ$, there is some change at higher twist angles. This may be due to effects in the moir\'e band structure, which becomes significantly more dispersive at higher twist angles \cite{wang_correlated_2020,devakul_magic_2021}. 

\subsection{Discussion of the relative measurement of $\mu$}

\begin{figure*}[h]
    \renewcommand{\thefigure}{S\arabic{figure}}
    \centering
    \includegraphics[scale=1.0]{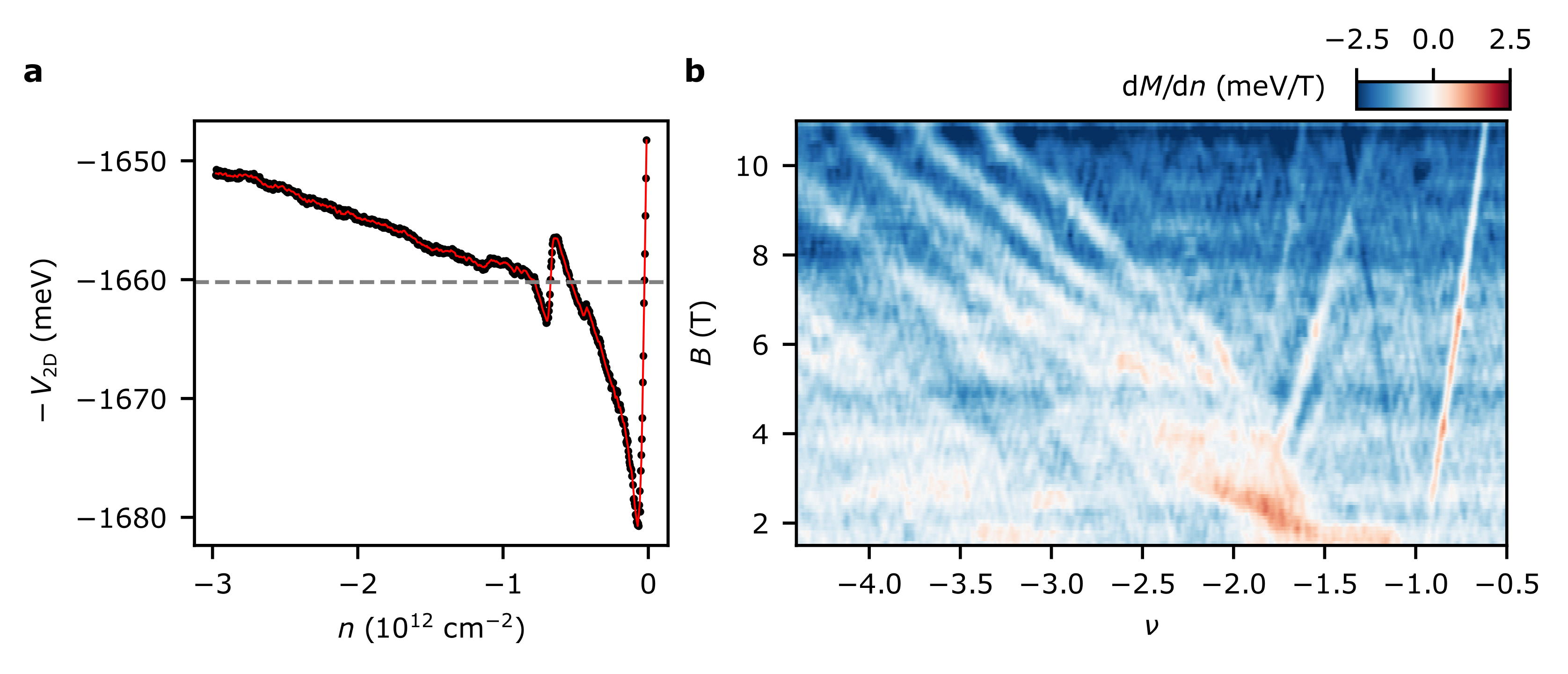}
    \caption{\textbf{Direct measurement of $\mu$.} \textbf{a}, $\mu$ measured as a function of $n$ at $B = 10.04$ T; chemical potential is given (up to a constant offset) by $-1 \times V_{\mathrm{2D}}$ where $V_{\mathrm{2D}}$ is the voltage applied to the tWSe$_2$ sample, feeding back to a single point on the Coulomb blockade oscillations. The red curve indicates smoothed data via a Savitzky-Golay filter. The dashed gray line indicates the measured chemical potential at the center of $\nu = -1$ gap, which we use a reference point for $\mu$ for data shown in the main text [i.e., setting $\mu(\nu = -1) = 0$]. \textbf{b}, d$M_z$/d$n$ identical to that shown in Fig.~2a in the main text but without the choice of setting $\mu(\nu = -1) = 0$. This shows a broad, negative background to d$M_z$/d$n$ that increases with increasing field, but features are otherwise unchanged.}
    \label{fig:SupMubgnd}
\end{figure*}

In the Methods, we describe how we extract the magnetization through a Maxwell relation. This depends on our measurement of $\mu(\nu,B)$, which is built up from a number of measurements of $\mu(\nu)$ at fixed $B$. In order to compare $\mu(\nu)$ measured at different magnetic fields, we choose a reference for each curve, which is $\mu(\nu = -1) = 0$ (Fig.~\ref{fig:SupMubgnd}). This choice of a reference for the chemical potential is common in order to avoid small drifts in the measurement that confound with the desired signal \cite{saito_isospin_2021,park_flavour_2021,rozen_entropic_2021}.

Without setting this convention, our measurement of $\mu(n)$ effectively measures the chemical potential relative to an energy level of the island of the SET. If we take derivatives of $\mu(n,B)$ in this way to extract d$M_z/$d$n$, we measure an overall broad negative background to d$M_z$/d$n$, of order of a few meV/T (Fig.~\ref{fig:SupMubgnd}b). This indicates that the voltage required to maintain a constant SET current drifts slightly in magnetic field. The sign of this signal is consistent with filling spin-polarized holes in this regime. However, we cannot rule out some influence from extrinsic effects on the energy levels within the SET or small physical drifts of the SET tip over the many hours required to take the measurement. Regardless of the origin of this background, the sign and relative magnitude of the main features discussed in the main text are unchanged. In measuring the chemical potential relative to $\nu = -1$, we remove this background, and any measurements of d$M_z$/d$n$ in the main text should strictly be interpreted as $\mathrm{d}M_z/\mathrm{d}n - \left(\mathrm{d}M_z/\mathrm{d}n \right)|_{\nu = -1}$.

